\documentclass[aps,prl,10pt,twocolumn,showpacs,superscriptaddress]{revtex4-1} 
\usepackage{graphicx}  % needed for figures
\usepackage{dcolumn}   % needed for some tables
\usepackage{array}
\usepackage{amssymb}
\usepackage{amsmath}
\usepackage{physics}
\usepackage{color}
\usepackage[dvipsnames]{xcolor}
\usepackage[colorlinks=true, citecolor={blue!80!black}, urlcolor={blue!50!black}, linkcolor = {blue!80!black}]{hyperref}

\DeclareMathOperator{\e}{e}

\begin{document}

\title{Suppressed Charge Dispersion via Resonant Tunneling \\ in a Single-Channel Transmon}

\author{A.~Kringh{\o}j}
\affiliation{Microsoft Quantum Lab Copenhagen and Center for Quantum Devices, Niels Bohr Institute, University of Copenhagen,
Universitetsparken 5, 2100 Copenhagen, Denmark}
\author{B.~van~Heck}
\affiliation{Microsoft Quantum, Station Q, University of California, Santa Barbara, California 93106-6105, USA}
%\affiliation{Microsoft Quantum Lab Copenhagen and Center for Quantum Devices, Niels Bohr Institute, University of Copenhagen,
%Universitetsparken 5, 2100 Copenhagen, Denmark}
\affiliation{Microsoft Quantum Lab Delft, Delft University of Technology, 2600 GA Delft, The Netherlands}
\author{T.~W.~Larsen}
\author{O.~Erlandsson}
\author{D.~Sabonis}
\affiliation{Microsoft Quantum Lab Copenhagen and Center for Quantum Devices, Niels Bohr Institute, University of Copenhagen,
Universitetsparken 5, 2100 Copenhagen, Denmark}
\author{P.~Krogstrup}
\affiliation{Microsoft Quantum Lab Copenhagen and Center for Quantum Devices, Niels Bohr Institute, University of Copenhagen,
Universitetsparken 5, 2100 Copenhagen, Denmark}
\affiliation{Microsoft Quantum Materials Lab Copenhagen, Kanalvej 7, 2800 Lyngby, Denmark}
\author{L.~Casparis}
\author{K.~D.~Petersson}
\author{C.~M.~Marcus}
\affiliation{Microsoft Quantum Lab Copenhagen and Center for Quantum Devices, Niels Bohr Institute, University of Copenhagen,
Universitetsparken 5, 2100 Copenhagen, Denmark}

\begin{abstract}
%We demonstrate strong suppression of charge dispersion in a semiconductor-based transmon qubit across Josephson resonances associated with a quantum dot in the junction. On resonance, dispersion is drastically reduced compared to conventional transmons with corresponding Josephson and charging energies. A model of qubit dispersion for a single-channel resonance is in quantitative agreement with experimental data.
We demonstrate strong suppression of charge dispersion in a semiconductor-based transmon qubit across Josephson resonances associated with a quantum dot in the junction. On resonance, dispersion is drastically reduced compared to conventional transmons with corresponding Josephson and charging energies. We develop a model of qubit dispersion for a single-channel resonance, which is in quantitative agreement with experimental data.

\end{abstract}

\maketitle

% Introduction

Superconducting circuits based on nonlinear Josephson junctions (JJ) form the basis of a broad array of coherent quantum devices used in applications ranging from radiation detectors to magnetometers to qubits~\cite{makhlin_2001, girvin_2014}. An important application is the transmon qubit, a variant of the Cooper pair box qubit~\cite{Bouchiat_1998} where the Josephson energy, $E_J$, of the junction exceeds the charging energy, $E_C = e^{2}/2C$, of the shunting capacitor with capacitance $C$. Designing qubits with ratio $E_{J}/E_{C}$ considerably greater than unity exponentially suppresses its charge character, correspondingly reducing its sensitivity to voltage noise and dramatically extending coherence~\cite{koch_2007, schreier_2008}. The tradeoff with increasing $E_{J}/E_{C}$ is reduced anharmonicity, which determines the minimal operation time due to leakage out of computational states~\cite{krantz_2019}.

The JJs used in superconducting qubits are almost exclusively based on superconductor-insulator-superconductor (SIS) tunnel junctions~\cite{paik_2011}, well described by a sinusoidal current-phase relation (CPR)~\cite{golubov_2004}. More recently, gate-voltage-tunable transmon qubits (gatemons) have been realized using superconductor-semiconductor-superconductor (S-Sm-S) JJs, where the Sm weak link was either a nanowire~\cite{larsen_2015, delange_2015}, a two-dimensional electron gas~\cite{Casparis_2018} or graphene~\cite{Kroll_2018, Wang_2019}. Such Sm weak links are typically quasiballistic, and, with Andreev processes~\cite{beenakker_1991} across the junction dominated by a small number of highly transmitting channels~\cite{Spanton_2017, Goffman_2017, Anharmonicity}. In this regime, the CPR is no longer sinusoidal, and anharmonicity deviates from the usual relations and tradeoffs involving $E_J$ and $E_C$~\cite{Anharmonicity}.

An expected consequence of large transmission among a few Andreev modes in the JJ is a suppression of the quantization of island charge, which vanishes entirely when the transmission of any mode reaches unity~\cite{flensberg_1993, matveev_1995, nazarov_1999}. Suppression of charge quantization in non-superconducting quantum dots has been well investigated experimentally~\cite{patel_1998, duncan_2000}, including a recent detailed study in a semiconductor quantum dot with vanishing level spacing due to an internal normal-metal contact~\cite{jezouin_2016}.
In a similar fashion, charge quantization on a JJ-coupled superconducting island is expected to be suppressed for highly transmissive modes and vanish for unity transmission of a mode~\cite{Averin1999}, irrespective of the ratio $E_J/E_C$, though to our knowledge this has not been previously investigated experimentally.

%device figure--------------------

\begin{figure}[!b]\vspace{-4mm} \includegraphics[width=1\columnwidth]{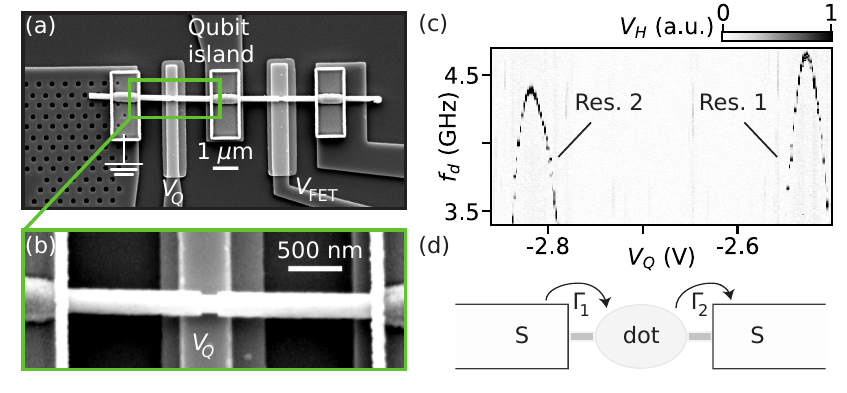}
    \caption{\vspace{-0mm} %Device geometry and spectroscopy close to pinch-off.
(a) Scanning electron micrograph (SEM) of the nanowire region of the qubit device. Two etched regions were formed (qubit junction and FET) controlled with bottom gates $V_Q$ and $V_\text{FET}$. (b) SEM of the qubit region highlighted (green square) in (a). (c) Two-tone spectroscopy measurements of the heterodyne transmission voltage $V_H$ at values of qubit gate voltage $V_Q$ just above complete depletion of the qubit junction and varying drive frequency $f_d$, yielding two resonances (Res.~1 and Res.~2) in the qubit frequency spectrum. (d) Sketch illustrating the principle of tunneling on and off a resonant dot level inside a Josephson junction connected to the superconducting leads by two tunnel barriers, characterized by tunnel rates $\Gamma_1$ and $\Gamma_2$.
    }
    \label{fig:device}
\end{figure}

%% in this letter

In this Letter, we investigate the charge dispersion in a nanowire-based gatemon qubit which shows strong suppression compared to a conventional metallic transmon qubit, when operated across resonances in the junction. As discussed below, resonances in the semiconductor JJ effectively bring the Andreev transmission of a single mode close to unity.
A comparison of experimental data to a simple model describing resonant Cooper pair transport across a single-mode junction~\cite{GlazmanMatveev1989,beenakkervanhouten,Devyatov1997,golubov_2004} yields striking agreement, supporting both the general feature of suppressed charge quantization at large transmission, and the additional feature that a dot resonance acts to provide an effective near-unity transmission of a single mode in a semiconductor JJ.

%device text --------------------

Measurements were performed on a gatemon qubit based on an InAs nanowire fully covered by 30\,nm epitaxial Al~\cite{krogstrup_2015}, as described previously~\cite{lead}.  Two $\sim150\,$nm segments of the Al shell were etched, forming gateable regions, as shown in Fig.~\ref{fig:device}(a), one serving as the qubit junction, controlled by gate voltage $V_Q$, and the other as a field-effect transistor (FET), allowing {\it in-situ} DC transport, controlled by $V_\text{FET}$~\cite{lead}.
All cQED measurements were carried out with the FET fully depleted ($V_\text{FET}=-3$\,V), so that the gatemon circuit consisted of one side of the qubit junction contacted to ground and the other to the capacitor island [Fig.~\ref{fig:device}(b)]. The island capacitance was designed to yield $E_C/h \sim500$\,MHz, allowing operation at intermediate $E_J/E_C \sim 1$0--20 so that charge dispersion was easily resolved.

Near the pinch-off voltage of the qubit junction ($V_{Q}\sim-3\,$V), the first visible features to appear in two-tone spectroscopy as $V_{Q}$ was tuned more positive were two narrow peaks in the qubit frequency, as shown in Fig.~\ref{fig:device}(c). We attribute these features to resonant tunneling of Cooper pairs through an accidental quantum dot formed in the junction [Fig.~\ref{fig:device}(d)], a common occurrence near full depletion \cite{chang_2015, Hart2019}. Corresponding resonant features were also observed in DC transport (FET opened) at similar values of $V_{Q}$~\cite{supplement}.

%% theory fig and text

\begin{figure}
    \centering
        \hspace{-2mm}\includegraphics[width=1\columnwidth]{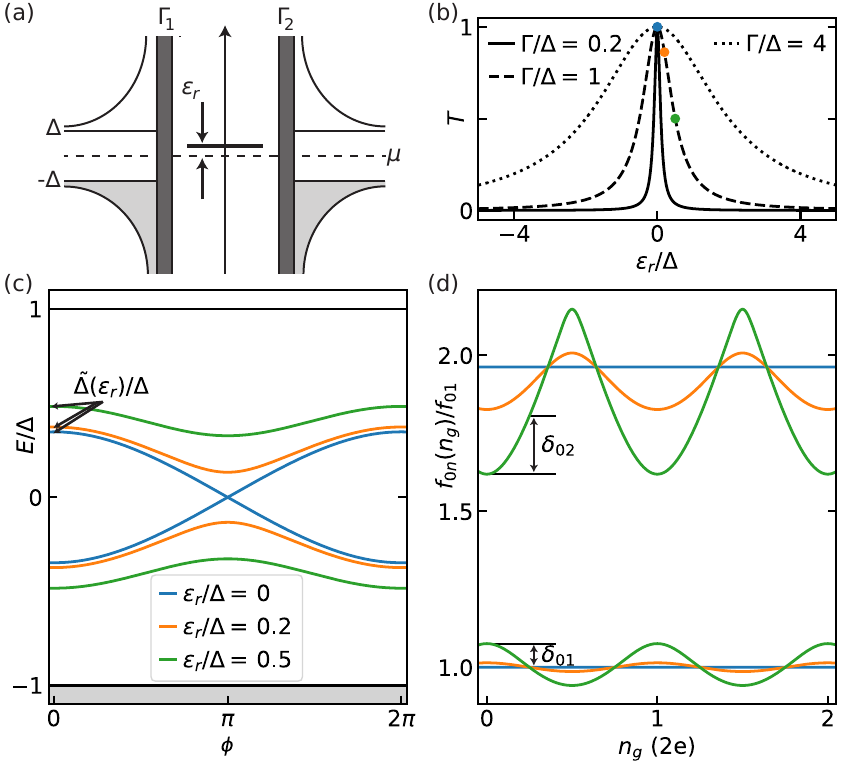}\vspace{-4mm}
    \caption{%Resonant tunneling model and qubit spectrum. 
    (a) Sketch of the energy density of states of a superconductor-dot-superconductor system. The superconductors are described by a standard BCS density of states with gap $\Delta$. A spin-degenerate level is located inside the JJ, detuned by $\epsilon_r$ from the Fermi level (dashed line). (b) Normal state transmission through the junction, $T$, as a function of $\epsilon_r$ for three different $\Gamma$ for $\Gamma_1=\Gamma_2$. Note that $T=1$ occurs for $\epsilon_r=0$ for all $\Gamma$. (c) Numerical solutions to Eq.~\eqref{eq:bound_state_equation} describing resonant tunneling for three different $\epsilon_r$ [coloured dots in (a)] and $\Gamma/\Delta=1$. The effective gap $\tilde{\Delta}(\epsilon_r)= E(0)$ (arrows) and continuum at $\pm E/\Delta=1$ (grey and white region) are indicated. (d) Numerical solutions to Eq.~\eqref{eq:model} showing two lowest transition frequencies $f_{01}(n_g)$ and $f_{02}(n_g)$ as a function of offset charge $n_g$. The frequencies are normalized to the $0\to 1$ degeneracy transition frequency $f_{01}(0.25)=f_{01}$ with dispersion amplitudes $\delta_{01}=f_{01}(0)-f_{01}(0.25)$ and $\delta_{02}=f_{02}(0.25)-f_{02}(0)$ indicated (arrows).
    }
    \label{fig:theory}\vspace{-4mm}
\end{figure}

To model the junction resonance, we consider a single spin-degenerate level at energy $\epsilon_r$, weakly coupled to the two superconducting leads via tunneling rates $\Gamma_1$ and $\Gamma_2$ [Fig.~\ref{fig:theory}(a)], and a Breit-Wigner form for the transmission~\cite{LarkinMatveev1987}, $T=4\Gamma_1\Gamma_2/(\epsilon_r^2 + \Gamma^2)$, where $\Gamma=\Gamma_1+\Gamma_2$. Transmission is maximal on resonance, $\epsilon_r=0$, where it reaches unity for symmetric barriers, $\Gamma_1=\Gamma_2$ [Fig.~\ref{fig:theory}(b)]. In the superconducting state, a pair of spin-degenerate Andreev bound states reside in the junction at energy $E$, given by~\cite{golubov_2004,beenakkervanhouten}
\begin{align}\label{eq:bound_state_equation}
2\sqrt{\Delta^2-E^2}\,E^2\,\Gamma + (\Delta^2-E^2)(E^2-\epsilon_r^2-\Gamma^2)&\nonumber \\
+\, 4\Delta^2\, \Gamma_1\Gamma_2\,\sin^2(\phi/2) &= 0
\end{align}
where $\Delta$ the superconducting gap and $\phi$ the phase difference across the junction~\cite{supplement}, as plotted in Fig.~\ref{fig:theory}(c).

The Andreev level spectrum consists of a spin-degenerate, phase-dependent bound state plus a continuum of quasiparticle states above the gap.
At $\phi=0$, the bound state energy $E(0)=\tilde\Delta$, varies between $\epsilon_r$ and $\Delta$ as $\Gamma$ increases~\cite{supplement}.
The energy gap at $\phi=\pi$ is proportional to the reflection amplitude $r=\sqrt{1-T}$ and thus vanishes at perfect transmission, yielding two decoupled $4\pi$-periodic branches.

We model the charging-energy-induced quantum fluctuations in $\phi$ via the Hamiltonian~\cite{Ivanov1998,Ivanov1999,Zazunov2005},
\begin{subequations}\label{eq:model}
\begin{align}
H&=4E_C\left(i\partial_\phi-n_g\right)^2+H_J\,,\\
H_J&=\tilde{\Delta}
\begin{bmatrix} 
\cos\left(\phi/2\right) & r\sin\left(\phi/2\right) \\r\sin\left(\phi/2\right) & -\cos\left(\phi/2\right)
\end{bmatrix}\,,
\end{align}
\end{subequations}
where $n_g$ is the charge induced on the island in units of $2e$.
The model above was originally derived for a superconducting quantum point contact~\cite{Ivanov1999}, and it is valid provided $E_C\ll \Delta$ and that the Andreev energies are well separated from the continuum.
The eigenvalues of $H_J$,
\begin{equation}
E = \pm \tilde{\Delta}[1-T\sin^2(\phi/2)]^{1/2},
\end{equation}
closely approximate the solutions of Eq.~\eqref{eq:bound_state_equation}~\cite{supplement}. 
We solve Eq.~\eqref{eq:model} numerically~\cite{supplement} to obtain the qubit energy levels $E_n$ as well as the associated transition frequencies $f_{nm}(n_g)=(E_m(n_g)-E_n(n_g))/h$ [Fig.~\ref{fig:theory}(d)].

A key feature of Eq.~\eqref{eq:model} is that it captures the Landau-Zener dynamics across the avoided crossing at $\phi=\pi$, which has a dramatic effect on charge dispersion of the qubit energy levels~\cite{Averin1999}.
Indeed, the charge dispersion is determined by the $2\pi$-tunneling amplitude of the phase below the Josephson potential energy barrier, which is suppressed by the probability of a diabatic passage to the excited branch of the Andreev spectrum.
This probability becomes large near perfect transmission, when $r\ll (E_C/\tilde\Delta)^{1/2}$.
At $r=0$, the $2\pi$-tunneling processes become forbidden, and the charge dispersion reaches a minimal value given by the amplitude for $4\pi$-tunneling~\footnote{Due to the large area below the potential barrier for $4\pi$-tunneling at perfect transmission, this residual dispersion can be estimated to be well below the experimentally achieved linewidth.}.
The remarkable flattening of the energy levels in this diabatic regime is illustrated in Fig.~\ref{fig:theory}(d).

%%Dispersion data and text --------------------------------

\begin{figure}
    \centering
        \hspace{-2mm}\includegraphics[width=1\columnwidth]{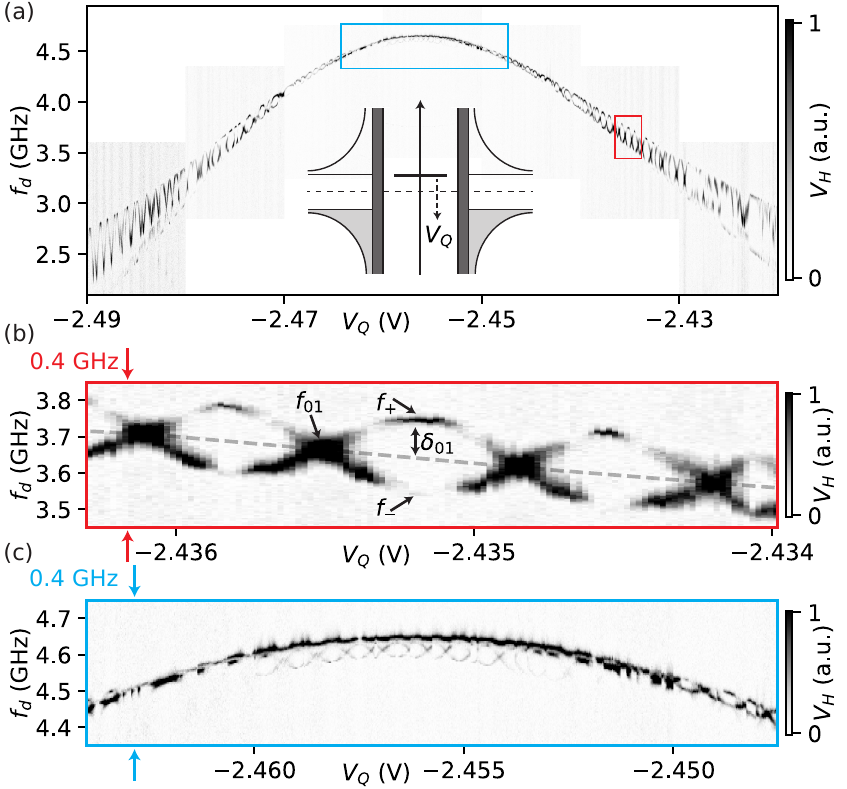}\vspace{-4mm}
    \caption{(a) Measurement of the heterodyne transmission voltage $V_H$ as a function of $V_Q$ and a varying qubit drive $f_d$ across one of two resonances (Res.~1). (Inset) Sketch of the energy density of states to illustrate the interpretation that $\epsilon_r$ is varied by $V_Q$. (b,~c) Zoom at the red (blue) region in (a) at the slope (peak) of the resonance spectrum. Note the same $f_d$ scale of 0.4\,GHz in both panels. Examples of maximal upper ($f_+$), minimal lower ($f_-$), and charge degeneracy ($f_{01}$) frequencies are indicated in (b) (single arrows). An example of the maximal charge dispersion amplitude $\delta_{01}=f_+-f_{01}$ is indicated (double arrow). Interpolated $f_{01}$ as a function of $V_Q$ is shown in (b) (grey dashed line).
    }
    \label{fig:rawdata}\vspace{-4mm}
\end{figure}

% ---------------------- experiment text

% measurement description
Measurements of charge dispersion across Res.~1 in Fig.~\ref{fig:device}(c) were carried out by finely sweeping $V_Q$ while performing two-tone spectroscopy using a rastered drive tone $f_d$ followed by a readout tone at $f_R\sim5.3\,$GHz [Fig.~\ref{fig:rawdata}(a)]. The fine sweep of $V_Q$ served two purposes; it both tuned the junction across the resonance and incremented the charge $n_g$ on the superconducting island, resulting in an oscillating pattern within a resonant envelope, appearing in the demodulated transmission voltage $V_H$~[Fig.~\ref{fig:rawdata}(a)]. The two counter-oscillating branches reflect fast quasiparticle poisoning of the island, which shifts the energy spectrum in Fig.~\ref{fig:theory}(d) by half a period ($1e$)~\cite{schreier_2008}.

% 01 dispersion data extraction

Qubit frequencies for both parity branches were extracted from the raw $V_H$ data using double Lorentzian fits for each $V_Q$, allowing determination of the maximal upper ($f_+$) and minimal lower ($f_-$) branch frequencies. At the charge degeneracy points a single Lorentzian fit was used to find $f_{01}$.
The charge dispersion amplitude, here defined $\delta_{01}=f_+-f_{01}$, was then extracted using an interpolated $f_{01}$ to determine  $f_+$ and $f_{01}$ at corresponding $V_Q$, as shown in Fig.~\ref{fig:rawdata}(b).
Near the top of the resonance, the two-photon transition frequency $f_{02}(n_g)/2$ was visible in the spectrum and overlaps with the lower frequency branch of the $f_{01}$ transition [Fig.~\ref{fig:rawdata}(c)]. As $\delta_{01}$ becomes comparable to the linewidth here we use the observed $f_{02}(n_g)/2$ to identify the $V_Q$ associated with charge degeneracy and maximal dispersion amplitude.
 
Measurements of charge dispersion across Res.~2 were done in a slightly different way. Rather than using $V_Q$ to span the resonance and vary $n_g$, for Res.~2, $n_g$ was varied by sweeping $V_\text{FET}$ (in the depleted regime) at fixed $V_{Q}$ giving roughly independent control of $\epsilon_{r}$ and $n_g$~\cite{supplement}. The observed behavior of Res.~1 and Res.~2 was the same.

% 01 dispersion Fig, fitresult and analysis ----------------------------
\begin{figure}
    \centering
        \hspace{-2mm}\includegraphics[width=1\columnwidth]{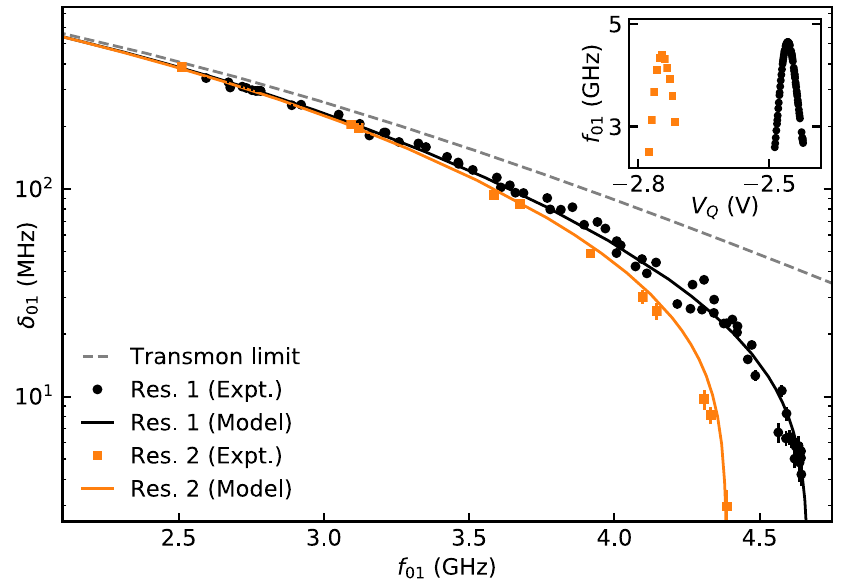}\vspace{-4mm}
    \caption{Extracted maximal dispersion amplitudes (black and orange data points) and fit results (black and orange curves) of the $0\to 1$ transition for both resonances (Res.~1 and Res.~2) as a function of qubit frequency $f_{01}$. The theory curves are fits of numerical solutions to Eq.~\eqref{eq:model} with fit parameters $E_C/h=539\,$MHz and $\Gamma/h=72\,(60)\,$GHz for Res.~1\,(2). Numerical $\delta_{01}$ (grey dashed line) for the standard transmon model with $E_C/h=539\,$MHz. Error bars are estimated from fit errors. (Inset) Extracted $f_{01}$ as a function of $V_Q$ for Res.~1 (black) and Res.~2 (orange).
    }
    \label{fig:01_result}\vspace{-4mm}
\end{figure}

% Description of the charge dispersion data
Figure~\ref{fig:01_result} shows a parametric plot of dispersion $\delta_{01}$ as a function of $f_{01}$ for both resonances, with the original dependence of $f_{01}$ on $V_Q$ shown in the inset.
As expected for transmons in general, $\delta_{01}$ decreases when $f_{01}$ increases due to an increase in $E_{J}$.
In the $f_{01} \lesssim 3.5$\,GHz range, corresponding to the tails of the two resonances, $\delta_{01}$ decays approximately exponentially as $f_{01}$ is increased.
However, for the $f_{01}\gtrsim 4$\,GHz range, near the top of the two resonances, we observe the onset of a sharper decrease towards vanishing $\delta_{01}$, strongly deviating from the exponential suppression expected in standard transmon qubits.

%Analysis of the data
To quantitatively compare the observed charge dispersion across the resonances to the model, Eq.~\eqref{eq:model}, we first fix $\Delta=190~\mu\text{eV}$ based on tunneling spectroscopy measurements at $V_\text{FET}=+4\,$V, where the FET is open~\cite{lead}.
For simplicity we take the tunnel barriers to be symmetric and only allow $V_Q$ to tune $\epsilon_r$. We then fit $E_C$ (the same for both resonances) and $\Gamma$ (allowed to be different for each resonance). Results are shown in Fig.~\ref{fig:01_result}, with $E_C/h=539\,$MHz (comparable to the electrostatic model~\cite{comsol} value $512$\,MHz) and $\Gamma/h=72$\,GHz for Res.~1, and $\Gamma/h=60$\,GHz for Res.~2.

Comparing $\delta_{01}$ to the prediction for a conventional transmon model based on the Hamiltonian $H_T=4E_C(n-n_g)^2-E_J\cos\phi$, for $E_C/h=539\,$MHz, highlights the suppressed dispersion observed experimentally and in the resonance model. The conventional model agrees with the experimental data and with the resonant level model only at low values of $f_{01}$, as expected for a decreasing transmission coefficient, where the sinusoidal CPR is recovered and the Landau-Zener dynamics becomes irrelevant.

When $V_Q$ is turned more positive, we no longer observed narrow, symmetric resonances associated with resonant tunneling. Instead, we observe a non-monotonic spectrum much less susceptible to changes in $V_Q$. In this regime, we also observe a deviation in the charge dispersion compared to the value predicted by $H_T$~\cite{supplement}. However, the suppression is not as pronounced as observed across the two resonances. 
We interpret this as crossing to a regime where the Andreev processes are no longer mediated by a resonant level and instead is described by a few gate tunable transmission coefficients~\cite{doh_2005, Spanton_2017, Goffman_2017, Anharmonicity}, not reaching values similarly close to unity.

% 02 result ----------------------------
\begin{figure}
    \centering
        \includegraphics[width=1\columnwidth]{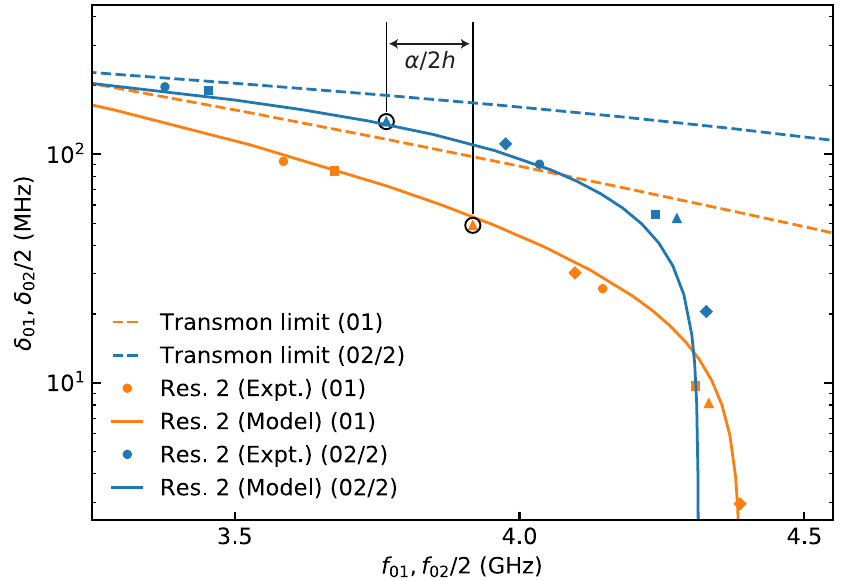}\vspace{-2mm}
    \caption{\vspace{-0mm} Extracted maximal dispersion amplitudes (orange and blue data points) and fit result (orange and blue curves) of the $0\to 1$ and $0\to 2$ transitions of Res.~2, respectively. The theory curves correspond to numerical solutions to Eq.~\eqref{eq:model} with $E_C/h=539\,$MHz and $\Gamma/h=60\,$GHz. Numerical $\delta_{01}$ (orange dashed line) and $\delta_{02}/2$ (blue dashed line) based on $H_T$ with $E_C/h=539\,$MHz. The frequency differences between corresponding pairs of data points taken at same $V_Q$ (matching shapes) are equal to $\alpha/2$, with one example indicated.
    }
    \label{fig:02_result}
\end{figure}

We also examine charge dispersion for the two-photon ($0\to 2$) transition frequencies of Res.~2. By increasing the power and repeating the scans used to extract $\delta_{01}$ we both excite the $0\to 1$ and the $0\to 2$ transitions.
We define the $0\to 2$ charge dispersion amplitude $\delta_{02}=f_{02}-f_{02,-}$, where $f_{02,-}$ and $f_{02}$ are the minimal lower branch and degeneracy frequency, respectively. This operative definition is chosen as the upper branch of the $0\to 2$ transition interferes with the lower branch of that of $0\to 1$.
Results for both $\delta_{01}$ and $\delta_{02}/2$ are shown in Fig.~\ref{fig:02_result}.
Both theory curves are obtained by solving Eq.~\eqref{eq:model} for the same parameters as in Fig.~\ref{fig:01_result}, again showing striking agreement between theory and experiment. We also compare the measured $\delta_{02}/2$ with numerical solutions to $H_T$, again yielding roughly an order of magnitude deviation at resonance~\footnote{Numerical code and data accompanying the analysis of Figs.~4 and 5 are found at:\\ \url{https://github.com/anderskringhoej/Dispersion}.}.
Finally, we emphasize that the finite frequency difference between the pairs of data points is equal to half the anharmonicity $\alpha$, as $f_{02}/2-f_{01}=1/2\left(f_{12}-f_{01}\right)=\alpha/2h$. This illustrates that $\delta_{0i}\rightarrow0$ can be achieved without $\alpha\rightarrow0$ and in principle for much larger $\alpha$.

% Model limitations,
Minor deviations between experiment and model may be attributed to effects of electron-electron interactions in the quantum dot, which are not included in the model~\cite{Vecino2003,MartinRodero2011,Hart2019} as well as fluctuations in the ratio $\Gamma_1/\Gamma_2$ as a function of $V_Q$. 

% Conclusions
In summary, we have observed and modeled the strong suppression of the charge dispersion in a single-channel transmon across a junction resonance, obtaining excellent agreement between experiment and theory. Our results suggest that charge dispersion can be suppressed without the necessity of large $E_J/E_C$ ratios. Future implementation of controlled dot structures or QPC junctions to controllably achieve transmissions near unity may be a path to engineer superconducting qubits with vanishing charge dispersion and large anharmonicity. Additionally a controllable near-unity junction would allow for deterministic tuning of the spectrum in Andreev qubits~\cite{janvier_2016, Hays_2018}. 
Similar results are presented in Ref.~\cite{arno}, in coordination with results reported here.
\let\oldaddcontentsline\addcontentsline% Store \addcontentsline
\renewcommand{\addcontentsline}[3]{}% Make \addcontentsline a no-op
\begin{acknowledgments}
This work was supported by Microsoft and the Danish National Research Foundation. 
We acknowledge discussions with Karsten Flensberg, Michael Hell, and Martin Leijnse that inspired the ideas of the experiment.
We thank Andrey Antipov, Arno Bargerbos, Gijs de Lange, Angela Kou, Roman Lutchyn, and Chaitanya Murthy for useful discussions. 
We acknowledge Marina Hesselberg, Karthik Jambunathan, Robert McNeil, Karolis Parfeniukas, Agnieszka Telecka, Shivendra Upadhyay, and Sachin Yadav for the device fabrication.
BvH thanks the Center for Quantum Devices, Niels Bohr Institute for the hospitality during part of the time in which this study was carried out. PK acknowledges funding from the European Research Commision through the grant agreement 'HEMs-DAM' no.716655.

\end{acknowledgments}
\let\addcontentsline\oldaddcontentsline% Restore \addcontentsline
%\bibliography{references}

\begin{thebibliography}{45}%
\makeatletter
\providecommand \@ifxundefined [1]{%
 \@ifx{#1\undefined}
}%
\providecommand \@ifnum [1]{%
 \ifnum #1\expandafter \@firstoftwo
 \else \expandafter \@secondoftwo
 \fi
}%
\providecommand \@ifx [1]{%
 \ifx #1\expandafter \@firstoftwo
 \else \expandafter \@secondoftwo
 \fi
}%
\providecommand \natexlab [1]{#1}%
\providecommand \enquote  [1]{``#1''}%
\providecommand \bibnamefont  [1]{#1}%
\providecommand \bibfnamefont [1]{#1}%
\providecommand \citenamefont [1]{#1}%
\providecommand \href@noop [0]{\@secondoftwo}%
\providecommand \href [0]{\begingroup \@sanitize@url \@href}%
\providecommand \@href[1]{\@@startlink{#1}\@@href}%
\providecommand \@@href[1]{\endgroup#1\@@endlink}%
\providecommand \@sanitize@url [0]{\catcode `\\12\catcode `\$12\catcode
  `\&12\catcode `\#12\catcode `\^12\catcode `\_12\catcode `\%12\relax}%
\providecommand \@@startlink[1]{}%
\providecommand \@@endlink[0]{}%
\providecommand \url  [0]{\begingroup\@sanitize@url \@url }%
\providecommand \@url [1]{\endgroup\@href {#1}{\urlprefix }}%
\providecommand \urlprefix  [0]{URL }%
\providecommand \Eprint [0]{\href }%
\providecommand \doibase [0]{http://dx.doi.org/}%
\providecommand \selectlanguage [0]{\@gobble}%
\providecommand \bibinfo  [0]{\@secondoftwo}%
\providecommand \bibfield  [0]{\@secondoftwo}%
\providecommand \translation [1]{[#1]}%
\providecommand \BibitemOpen [0]{}%
\providecommand \bibitemStop [0]{}%
\providecommand \bibitemNoStop [0]{.\EOS\space}%
\providecommand \EOS [0]{\spacefactor3000\relax}%
\providecommand \BibitemShut  [1]{\csname bibitem#1\endcsname}%
\let\auto@bib@innerbib\@empty
%</preamble>
\bibitem [{\citenamefont {Makhlin}\ \emph {et~al.}(2001)\citenamefont
  {Makhlin}, \citenamefont {Sch\"on},\ and\ \citenamefont
  {Shnirman}}]{makhlin_2001}%
  \BibitemOpen
  \bibfield  {author} {\bibinfo {author} {\bibfnamefont {Y.}~\bibnamefont
  {Makhlin}}, \bibinfo {author} {\bibfnamefont {G.}~\bibnamefont {Sch\"on}}, \
  and\ \bibinfo {author} {\bibfnamefont {A.}~\bibnamefont {Shnirman}},\ }\href
  {\doibase 10.1103/RevModPhys.73.357} {\bibfield  {journal} {\bibinfo
  {journal} {Rev. Mod. Phys.}\ }\textbf {\bibinfo {volume} {73}},\ \bibinfo
  {pages} {357} (\bibinfo {year} {2001})}\BibitemShut {NoStop}%
\bibitem [{\citenamefont {Girvin}(2014)}]{girvin_2014}%
  \BibitemOpen
  \bibfield  {author} {\bibinfo {author} {\bibfnamefont {S.~M.}\ \bibnamefont
  {Girvin}},\ }\href@noop {} {\emph {\bibinfo {title} {\textrm{in} Quantum
  Machines: Measurement and Control of Engineered Quantum Systems}}},\ Lecture
  Notes of the Les Houches Summer School: Volume 96, July 2011\ (\bibinfo
  {publisher} {Oxford University Press},\ \bibinfo {year} {2014})\BibitemShut
  {NoStop}%
\bibitem [{\citenamefont {Bouchiat}\ \emph {et~al.}(1998)\citenamefont
  {Bouchiat}, \citenamefont {Vion}, \citenamefont {Joyez}, \citenamefont
  {Esteve},\ and\ \citenamefont {Devoret}}]{Bouchiat_1998}%
  \BibitemOpen
  \bibfield  {author} {\bibinfo {author} {\bibfnamefont {V.}~\bibnamefont
  {Bouchiat}}, \bibinfo {author} {\bibfnamefont {D.}~\bibnamefont {Vion}},
  \bibinfo {author} {\bibfnamefont {P.}~\bibnamefont {Joyez}}, \bibinfo
  {author} {\bibfnamefont {D.}~\bibnamefont {Esteve}}, \ and\ \bibinfo {author}
  {\bibfnamefont {M.~H.}\ \bibnamefont {Devoret}},\ }\href {\doibase
  10.1238/physica.topical.076a00165} {\bibfield  {journal} {\bibinfo  {journal}
  {Physica Scripta}\ }\textbf {\bibinfo {volume} {T76}},\ \bibinfo {pages}
  {165} (\bibinfo {year} {1998})}\BibitemShut {NoStop}%
\bibitem [{\citenamefont {Koch}\ \emph {et~al.}(2007)\citenamefont {Koch},
  \citenamefont {Yu}, \citenamefont {Gambetta}, \citenamefont {Houck},
  \citenamefont {Schuster}, \citenamefont {Majer}, \citenamefont {Blais},
  \citenamefont {Devoret}, \citenamefont {Girvin},\ and\ \citenamefont
  {Schoelkopf}}]{koch_2007}%
  \BibitemOpen
  \bibfield  {author} {\bibinfo {author} {\bibfnamefont {J.}~\bibnamefont
  {Koch}}, \bibinfo {author} {\bibfnamefont {T.~M.}\ \bibnamefont {Yu}},
  \bibinfo {author} {\bibfnamefont {J.~M.}\ \bibnamefont {Gambetta}}, \bibinfo
  {author} {\bibfnamefont {A.~A.}\ \bibnamefont {Houck}}, \bibinfo {author}
  {\bibfnamefont {D.~I.}\ \bibnamefont {Schuster}}, \bibinfo {author}
  {\bibfnamefont {J.}~\bibnamefont {Majer}}, \bibinfo {author} {\bibfnamefont
  {A.}~\bibnamefont {Blais}}, \bibinfo {author} {\bibfnamefont {M.~H.}\
  \bibnamefont {Devoret}}, \bibinfo {author} {\bibfnamefont {S.~M.}\
  \bibnamefont {Girvin}}, \ and\ \bibinfo {author} {\bibfnamefont {R.~J.}\
  \bibnamefont {Schoelkopf}},\ }\href {\doibase 10.1103/PhysRevA.76.042319}
  {\bibfield  {journal} {\bibinfo  {journal} {Phys. Rev. A}\ }\textbf {\bibinfo
  {volume} {76}},\ \bibinfo {pages} {042319} (\bibinfo {year}
  {2007})}\BibitemShut {NoStop}%
\bibitem [{\citenamefont {Schreier}\ \emph {et~al.}(2008)\citenamefont
  {Schreier}, \citenamefont {Houck}, \citenamefont {Koch}, \citenamefont
  {Schuster}, \citenamefont {Johnson}, \citenamefont {Chow}, \citenamefont
  {Gambetta}, \citenamefont {Majer}, \citenamefont {Frunzio}, \citenamefont
  {Devoret}, \citenamefont {Girvin},\ and\ \citenamefont
  {Schoelkopf}}]{schreier_2008}%
  \BibitemOpen
  \bibfield  {author} {\bibinfo {author} {\bibfnamefont {J.~A.}\ \bibnamefont
  {Schreier}}, \bibinfo {author} {\bibfnamefont {A.~A.}\ \bibnamefont {Houck}},
  \bibinfo {author} {\bibfnamefont {J.}~\bibnamefont {Koch}}, \bibinfo {author}
  {\bibfnamefont {D.~I.}\ \bibnamefont {Schuster}}, \bibinfo {author}
  {\bibfnamefont {B.~R.}\ \bibnamefont {Johnson}}, \bibinfo {author}
  {\bibfnamefont {J.~M.}\ \bibnamefont {Chow}}, \bibinfo {author}
  {\bibfnamefont {J.~M.}\ \bibnamefont {Gambetta}}, \bibinfo {author}
  {\bibfnamefont {J.}~\bibnamefont {Majer}}, \bibinfo {author} {\bibfnamefont
  {L.}~\bibnamefont {Frunzio}}, \bibinfo {author} {\bibfnamefont {M.~H.}\
  \bibnamefont {Devoret}}, \bibinfo {author} {\bibfnamefont {S.~M.}\
  \bibnamefont {Girvin}}, \ and\ \bibinfo {author} {\bibfnamefont {R.~J.}\
  \bibnamefont {Schoelkopf}},\ }\href {\doibase 10.1103/PhysRevB.77.180502}
  {\bibfield  {journal} {\bibinfo  {journal} {Phys. Rev. B}\ }\textbf {\bibinfo
  {volume} {77}},\ \bibinfo {pages} {180502(R)} (\bibinfo {year}
  {2008})}\BibitemShut {NoStop}%
\bibitem [{\citenamefont {Krantz}\ \emph {et~al.}(2019)\citenamefont {Krantz},
  \citenamefont {Kjaergaard}, \citenamefont {Yan}, \citenamefont {Orlando},
  \citenamefont {Gustavsson},\ and\ \citenamefont {Oliver}}]{krantz_2019}%
  \BibitemOpen
  \bibfield  {author} {\bibinfo {author} {\bibfnamefont {P.}~\bibnamefont
  {Krantz}}, \bibinfo {author} {\bibfnamefont {M.}~\bibnamefont {Kjaergaard}},
  \bibinfo {author} {\bibfnamefont {F.}~\bibnamefont {Yan}}, \bibinfo {author}
  {\bibfnamefont {T.~P.}\ \bibnamefont {Orlando}}, \bibinfo {author}
  {\bibfnamefont {S.}~\bibnamefont {Gustavsson}}, \ and\ \bibinfo {author}
  {\bibfnamefont {W.~D.}\ \bibnamefont {Oliver}},\ }\href@noop {} {\bibfield
  {journal} {\bibinfo  {journal} {Appl. Phys. Rev.}\ }\textbf {\bibinfo
  {volume} {6}},\ \bibinfo {pages} {021318} (\bibinfo {year}
  {2019})}\BibitemShut {NoStop}%
\bibitem [{\citenamefont {Paik}\ \emph {et~al.}(2011)\citenamefont {Paik},
  \citenamefont {Schuster}, \citenamefont {Bishop}, \citenamefont {Kirchmair},
  \citenamefont {Catelani}, \citenamefont {Sears}, \citenamefont {Johnson},
  \citenamefont {Reagor}, \citenamefont {Frunzio}, \citenamefont {Glazman},
  \citenamefont {Girvin}, \citenamefont {Devoret},\ and\ \citenamefont
  {Schoelkopf}}]{paik_2011}%
  \BibitemOpen
  \bibfield  {author} {\bibinfo {author} {\bibfnamefont {H.}~\bibnamefont
  {Paik}}, \bibinfo {author} {\bibfnamefont {D.~I.}\ \bibnamefont {Schuster}},
  \bibinfo {author} {\bibfnamefont {L.~S.}\ \bibnamefont {Bishop}}, \bibinfo
  {author} {\bibfnamefont {G.}~\bibnamefont {Kirchmair}}, \bibinfo {author}
  {\bibfnamefont {G.}~\bibnamefont {Catelani}}, \bibinfo {author}
  {\bibfnamefont {A.~P.}\ \bibnamefont {Sears}}, \bibinfo {author}
  {\bibfnamefont {B.~R.}\ \bibnamefont {Johnson}}, \bibinfo {author}
  {\bibfnamefont {M.~J.}\ \bibnamefont {Reagor}}, \bibinfo {author}
  {\bibfnamefont {L.}~\bibnamefont {Frunzio}}, \bibinfo {author} {\bibfnamefont
  {L.~I.}\ \bibnamefont {Glazman}}, \bibinfo {author} {\bibfnamefont {S.~M.}\
  \bibnamefont {Girvin}}, \bibinfo {author} {\bibfnamefont {M.~H.}\
  \bibnamefont {Devoret}}, \ and\ \bibinfo {author} {\bibfnamefont {R.~J.}\
  \bibnamefont {Schoelkopf}},\ }\href@noop {} {\bibfield  {journal} {\bibinfo
  {journal} {Phys. Rev. Lett.}\ }\textbf {\bibinfo {volume} {107}},\ \bibinfo
  {pages} {240501} (\bibinfo {year} {2011})}\BibitemShut {NoStop}%
\bibitem [{\citenamefont {Golubov}\ \emph {et~al.}(2004)\citenamefont
  {Golubov}, \citenamefont {Kupriyanov},\ and\ \citenamefont
  {Il'ichev}}]{golubov_2004}%
  \BibitemOpen
  \bibfield  {author} {\bibinfo {author} {\bibfnamefont {A.~A.}\ \bibnamefont
  {Golubov}}, \bibinfo {author} {\bibfnamefont {M.~Y.}\ \bibnamefont
  {Kupriyanov}}, \ and\ \bibinfo {author} {\bibfnamefont {E.}~\bibnamefont
  {Il'ichev}},\ }\href {\doibase 10.1103/RevModPhys.76.411} {\bibfield
  {journal} {\bibinfo  {journal} {Rev. Mod. Phys.}\ }\textbf {\bibinfo {volume}
  {76}},\ \bibinfo {pages} {411} (\bibinfo {year} {2004})}\BibitemShut
  {NoStop}%
\bibitem [{\citenamefont {Larsen}\ \emph {et~al.}(2015)\citenamefont {Larsen},
  \citenamefont {Petersson}, \citenamefont {Kuemmeth}, \citenamefont
  {Jespersen}, \citenamefont {Krogstrup}, \citenamefont {Nygard},\ and\
  \citenamefont {Marcus}}]{larsen_2015}%
  \BibitemOpen
  \bibfield  {author} {\bibinfo {author} {\bibfnamefont {T.~W.}\ \bibnamefont
  {Larsen}}, \bibinfo {author} {\bibfnamefont {K.~D.}\ \bibnamefont
  {Petersson}}, \bibinfo {author} {\bibfnamefont {F.}~\bibnamefont {Kuemmeth}},
  \bibinfo {author} {\bibfnamefont {T.~S.}\ \bibnamefont {Jespersen}}, \bibinfo
  {author} {\bibfnamefont {P.}~\bibnamefont {Krogstrup}}, \bibinfo {author}
  {\bibfnamefont {J.}~\bibnamefont {Nygard}}, \ and\ \bibinfo {author}
  {\bibfnamefont {C.~M.}\ \bibnamefont {Marcus}},\ }\href@noop {} {\bibfield
  {journal} {\bibinfo  {journal} {Phys. Rev. Lett.}\ }\textbf {\bibinfo
  {volume} {115}},\ \bibinfo {pages} {127001} (\bibinfo {year}
  {2015})}\BibitemShut {NoStop}%
\bibitem [{\citenamefont {de~Lange}\ \emph {et~al.}(2015)\citenamefont
  {de~Lange}, \citenamefont {van Heck}, \citenamefont {Bruno}, \citenamefont
  {van Woerkom}, \citenamefont {Geresdi}, \citenamefont {Plissard},
  \citenamefont {Bakkers}, \citenamefont {Akhmerov},\ and\ \citenamefont
  {DiCarlo}}]{delange_2015}%
  \BibitemOpen
  \bibfield  {author} {\bibinfo {author} {\bibfnamefont {G.}~\bibnamefont
  {de~Lange}}, \bibinfo {author} {\bibfnamefont {B.}~\bibnamefont {van Heck}},
  \bibinfo {author} {\bibfnamefont {A.}~\bibnamefont {Bruno}}, \bibinfo
  {author} {\bibfnamefont {D.~J.}\ \bibnamefont {van Woerkom}}, \bibinfo
  {author} {\bibfnamefont {A.}~\bibnamefont {Geresdi}}, \bibinfo {author}
  {\bibfnamefont {S.~R.}\ \bibnamefont {Plissard}}, \bibinfo {author}
  {\bibfnamefont {E.~P. A.~M.}\ \bibnamefont {Bakkers}}, \bibinfo {author}
  {\bibfnamefont {A.~R.}\ \bibnamefont {Akhmerov}}, \ and\ \bibinfo {author}
  {\bibfnamefont {L.}~\bibnamefont {DiCarlo}},\ }\href@noop {} {\bibfield
  {journal} {\bibinfo  {journal} {Phys. Rev. Lett.}\ }\textbf {\bibinfo
  {volume} {115}},\ \bibinfo {pages} {127002} (\bibinfo {year}
  {2015})}\BibitemShut {NoStop}%
\bibitem [{\citenamefont {Casparis}\ \emph {et~al.}(2018)\citenamefont
  {Casparis}, \citenamefont {Connolly}, \citenamefont {Kjaergaard},
  \citenamefont {Pearson}, \citenamefont {Kringh{\o}j}, \citenamefont {Larsen},
  \citenamefont {Kuemmeth}, \citenamefont {Wang}, \citenamefont {Thomas},
  \citenamefont {Gronin}, \citenamefont {Gardner}, \citenamefont {Manfra},
  \citenamefont {Marcus},\ and\ \citenamefont {Petersson}}]{Casparis_2018}%
  \BibitemOpen
  \bibfield  {author} {\bibinfo {author} {\bibfnamefont {L.}~\bibnamefont
  {Casparis}}, \bibinfo {author} {\bibfnamefont {M.~R.}\ \bibnamefont
  {Connolly}}, \bibinfo {author} {\bibfnamefont {M.}~\bibnamefont
  {Kjaergaard}}, \bibinfo {author} {\bibfnamefont {N.~J.}\ \bibnamefont
  {Pearson}}, \bibinfo {author} {\bibfnamefont {A.}~\bibnamefont
  {Kringh{\o}j}}, \bibinfo {author} {\bibfnamefont {T.~W.}\ \bibnamefont
  {Larsen}}, \bibinfo {author} {\bibfnamefont {F.}~\bibnamefont {Kuemmeth}},
  \bibinfo {author} {\bibfnamefont {T.}~\bibnamefont {Wang}}, \bibinfo {author}
  {\bibfnamefont {C.}~\bibnamefont {Thomas}}, \bibinfo {author} {\bibfnamefont
  {S.}~\bibnamefont {Gronin}}, \bibinfo {author} {\bibfnamefont {G.~C.}\
  \bibnamefont {Gardner}}, \bibinfo {author} {\bibfnamefont {M.~J.}\
  \bibnamefont {Manfra}}, \bibinfo {author} {\bibfnamefont {C.~M.}\
  \bibnamefont {Marcus}}, \ and\ \bibinfo {author} {\bibfnamefont {K.~D.}\
  \bibnamefont {Petersson}},\ }\href@noop {} {\bibfield  {journal} {\bibinfo
  {journal} {Nat. Nanotechnol.}\ }\textbf {\bibinfo {volume} {13}},\ \bibinfo
  {pages} {915} (\bibinfo {year} {2018})}\BibitemShut {NoStop}%
\bibitem [{\citenamefont {Kroll}\ \emph {et~al.}(2018)\citenamefont {Kroll},
  \citenamefont {Uilhoorn}, \citenamefont {van~der Enden}, \citenamefont
  {de~Jong}, \citenamefont {Watanabe}, \citenamefont {Taniguchi}, \citenamefont
  {Goswami}, \citenamefont {Cassidy},\ and\ \citenamefont
  {Kouwenhoven}}]{Kroll_2018}%
  \BibitemOpen
  \bibfield  {author} {\bibinfo {author} {\bibfnamefont {J.}~\bibnamefont
  {Kroll}}, \bibinfo {author} {\bibfnamefont {W.}~\bibnamefont {Uilhoorn}},
  \bibinfo {author} {\bibfnamefont {K.}~\bibnamefont {van~der Enden}}, \bibinfo
  {author} {\bibfnamefont {D.}~\bibnamefont {de~Jong}}, \bibinfo {author}
  {\bibfnamefont {K.}~\bibnamefont {Watanabe}}, \bibinfo {author}
  {\bibfnamefont {T.}~\bibnamefont {Taniguchi}}, \bibinfo {author}
  {\bibfnamefont {S.}~\bibnamefont {Goswami}}, \bibinfo {author} {\bibfnamefont
  {M.}~\bibnamefont {Cassidy}}, \ and\ \bibinfo {author} {\bibfnamefont
  {L.}~\bibnamefont {Kouwenhoven}},\ }\href@noop {} {\bibfield  {journal}
  {\bibinfo  {journal} {Nat. Commun.}\ }\textbf {\bibinfo {volume} {9}},\
  \bibinfo {pages} {4615} (\bibinfo {year} {2018})}\BibitemShut {NoStop}%
\bibitem [{\citenamefont {Wang}\ \emph {et~al.}(2019)\citenamefont {Wang},
  \citenamefont {Rodan-Legrain}, \citenamefont {Bretheau}, \citenamefont
  {Campbell}, \citenamefont {Kannan}, \citenamefont {Kim}, \citenamefont
  {Kjaergaard}, \citenamefont {Krantz}, \citenamefont {Samach} \emph
  {et~al.}}]{Wang_2019}%
  \BibitemOpen
  \bibfield  {author} {\bibinfo {author} {\bibfnamefont {J.~I.-J.}\
  \bibnamefont {Wang}}, \bibinfo {author} {\bibfnamefont {D.}~\bibnamefont
  {Rodan-Legrain}}, \bibinfo {author} {\bibfnamefont {L.}~\bibnamefont
  {Bretheau}}, \bibinfo {author} {\bibfnamefont {D.~L.}\ \bibnamefont
  {Campbell}}, \bibinfo {author} {\bibfnamefont {B.}~\bibnamefont {Kannan}},
  \bibinfo {author} {\bibfnamefont {D.}~\bibnamefont {Kim}}, \bibinfo {author}
  {\bibfnamefont {M.}~\bibnamefont {Kjaergaard}}, \bibinfo {author}
  {\bibfnamefont {P.}~\bibnamefont {Krantz}}, \bibinfo {author} {\bibfnamefont
  {G.~O.}\ \bibnamefont {Samach}},  \emph {et~al.},\ }\href@noop {} {\bibfield
  {journal} {\bibinfo  {journal} {Nat. Nanotechnol.}\ }\textbf {\bibinfo
  {volume} {14}},\ \bibinfo {pages} {120} (\bibinfo {year} {2019})}\BibitemShut
  {NoStop}%
\bibitem [{\citenamefont {Beenakker}(1991)}]{beenakker_1991}%
  \BibitemOpen
  \bibfield  {author} {\bibinfo {author} {\bibfnamefont {C.~W.~J.}\
  \bibnamefont {Beenakker}},\ }\href {\doibase 10.1103/PhysRevLett.67.3836}
  {\bibfield  {journal} {\bibinfo  {journal} {Phys. Rev. Lett.}\ }\textbf
  {\bibinfo {volume} {67}},\ \bibinfo {pages} {3836} (\bibinfo {year}
  {1991})}\BibitemShut {NoStop}%
\bibitem [{\citenamefont {Spanton}\ \emph {et~al.}(2017)\citenamefont
  {Spanton}, \citenamefont {Deng}, \citenamefont {Vaitiek{\.e}nas},
  \citenamefont {Krogstrup}, \citenamefont {Nyg{\aa}rd}, \citenamefont
  {Marcus},\ and\ \citenamefont {Moler}}]{Spanton_2017}%
  \BibitemOpen
  \bibfield  {author} {\bibinfo {author} {\bibfnamefont {E.~M.}\ \bibnamefont
  {Spanton}}, \bibinfo {author} {\bibfnamefont {M.~T.}\ \bibnamefont {Deng}},
  \bibinfo {author} {\bibfnamefont {S.}~\bibnamefont {Vaitiek{\.e}nas}},
  \bibinfo {author} {\bibfnamefont {P.}~\bibnamefont {Krogstrup}}, \bibinfo
  {author} {\bibfnamefont {J.}~\bibnamefont {Nyg{\aa}rd}}, \bibinfo {author}
  {\bibfnamefont {C.~M.}\ \bibnamefont {Marcus}}, \ and\ \bibinfo {author}
  {\bibfnamefont {K.~A.}\ \bibnamefont {Moler}},\ }\href@noop {} {\bibfield
  {journal} {\bibinfo  {journal} {Nat. Phys.}\ }\textbf {\bibinfo {volume}
  {13}},\ \bibinfo {pages} {1177} (\bibinfo {year} {2017})}\BibitemShut
  {NoStop}%
\bibitem [{\citenamefont {Goffman}\ \emph {et~al.}(2017)\citenamefont
  {Goffman}, \citenamefont {Urbina}, \citenamefont {Pothier}, \citenamefont
  {Nyg{\aa}rd}, \citenamefont {Marcus},\ and\ \citenamefont
  {Krogstrup}}]{Goffman_2017}%
  \BibitemOpen
  \bibfield  {author} {\bibinfo {author} {\bibfnamefont {M.~F.}\ \bibnamefont
  {Goffman}}, \bibinfo {author} {\bibfnamefont {C.}~\bibnamefont {Urbina}},
  \bibinfo {author} {\bibfnamefont {H.}~\bibnamefont {Pothier}}, \bibinfo
  {author} {\bibfnamefont {J.}~\bibnamefont {Nyg{\aa}rd}}, \bibinfo {author}
  {\bibfnamefont {C.~M.}\ \bibnamefont {Marcus}}, \ and\ \bibinfo {author}
  {\bibfnamefont {P.}~\bibnamefont {Krogstrup}},\ }\href@noop {} {\bibfield
  {journal} {\bibinfo  {journal} {New Journal of Physics}\ }\textbf {\bibinfo
  {volume} {19}},\ \bibinfo {pages} {092002} (\bibinfo {year}
  {2017})}\BibitemShut {NoStop}%
\bibitem [{\citenamefont {Kringh\o{}j}\ \emph {et~al.}(2018)\citenamefont
  {Kringh\o{}j}, \citenamefont {Casparis}, \citenamefont {Hell}, \citenamefont
  {Larsen}, \citenamefont {Kuemmeth}, \citenamefont {Leijnse}, \citenamefont
  {Flensberg}, \citenamefont {Krogstrup}, \citenamefont {Nyg\aa{}rd},
  \citenamefont {Petersson},\ and\ \citenamefont {Marcus}}]{Anharmonicity}%
  \BibitemOpen
  \bibfield  {author} {\bibinfo {author} {\bibfnamefont {A.}~\bibnamefont
  {Kringh\o{}j}}, \bibinfo {author} {\bibfnamefont {L.}~\bibnamefont
  {Casparis}}, \bibinfo {author} {\bibfnamefont {M.}~\bibnamefont {Hell}},
  \bibinfo {author} {\bibfnamefont {T.~W.}\ \bibnamefont {Larsen}}, \bibinfo
  {author} {\bibfnamefont {F.}~\bibnamefont {Kuemmeth}}, \bibinfo {author}
  {\bibfnamefont {M.}~\bibnamefont {Leijnse}}, \bibinfo {author} {\bibfnamefont
  {K.}~\bibnamefont {Flensberg}}, \bibinfo {author} {\bibfnamefont
  {P.}~\bibnamefont {Krogstrup}}, \bibinfo {author} {\bibfnamefont
  {J.}~\bibnamefont {Nyg\aa{}rd}}, \bibinfo {author} {\bibfnamefont {K.~D.}\
  \bibnamefont {Petersson}}, \ and\ \bibinfo {author} {\bibfnamefont {C.~M.}\
  \bibnamefont {Marcus}},\ }\href {\doibase 10.1103/PhysRevB.97.060508}
  {\bibfield  {journal} {\bibinfo  {journal} {Phys. Rev. B}\ }\textbf {\bibinfo
  {volume} {97}},\ \bibinfo {pages} {060508(R)} (\bibinfo {year}
  {2018})}\BibitemShut {NoStop}%
\bibitem [{\citenamefont {Flensberg}(1993)}]{flensberg_1993}%
  \BibitemOpen
  \bibfield  {author} {\bibinfo {author} {\bibfnamefont {K.}~\bibnamefont
  {Flensberg}},\ }\href {\doibase 10.1103/PhysRevB.48.11156} {\bibfield
  {journal} {\bibinfo  {journal} {Phys. Rev. B}\ }\textbf {\bibinfo {volume}
  {48}},\ \bibinfo {pages} {11156} (\bibinfo {year} {1993})}\BibitemShut
  {NoStop}%
\bibitem [{\citenamefont {Matveev}(1995)}]{matveev_1995}%
  \BibitemOpen
  \bibfield  {author} {\bibinfo {author} {\bibfnamefont {K.~A.}\ \bibnamefont
  {Matveev}},\ }\href {\doibase 10.1103/PhysRevB.51.1743} {\bibfield  {journal}
  {\bibinfo  {journal} {Phys. Rev. B}\ }\textbf {\bibinfo {volume} {51}},\
  \bibinfo {pages} {1743} (\bibinfo {year} {1995})}\BibitemShut {NoStop}%
\bibitem [{\citenamefont {Nazarov}(1999)}]{nazarov_1999}%
  \BibitemOpen
  \bibfield  {author} {\bibinfo {author} {\bibfnamefont {Y.~V.}\ \bibnamefont
  {Nazarov}},\ }\href {\doibase 10.1103/PhysRevLett.82.1245} {\bibfield
  {journal} {\bibinfo  {journal} {Phys. Rev. Lett.}\ }\textbf {\bibinfo
  {volume} {82}},\ \bibinfo {pages} {1245} (\bibinfo {year}
  {1999})}\BibitemShut {NoStop}%
\bibitem [{\citenamefont {Patel}\ \emph {et~al.}(1998)\citenamefont {Patel},
  \citenamefont {Cronenwett}, \citenamefont {Stewart}, \citenamefont {Huibers},
  \citenamefont {Marcus}, \citenamefont {Duru\"oz}, \citenamefont {Harris},
  \citenamefont {Campman},\ and\ \citenamefont {Gossard}}]{patel_1998}%
  \BibitemOpen
  \bibfield  {author} {\bibinfo {author} {\bibfnamefont {S.~R.}\ \bibnamefont
  {Patel}}, \bibinfo {author} {\bibfnamefont {S.~M.}\ \bibnamefont
  {Cronenwett}}, \bibinfo {author} {\bibfnamefont {D.~R.}\ \bibnamefont
  {Stewart}}, \bibinfo {author} {\bibfnamefont {A.~G.}\ \bibnamefont
  {Huibers}}, \bibinfo {author} {\bibfnamefont {C.~M.}\ \bibnamefont {Marcus}},
  \bibinfo {author} {\bibfnamefont {C.~I.}\ \bibnamefont {Duru\"oz}}, \bibinfo
  {author} {\bibfnamefont {J.~S.}\ \bibnamefont {Harris}}, \bibinfo {author}
  {\bibfnamefont {K.}~\bibnamefont {Campman}}, \ and\ \bibinfo {author}
  {\bibfnamefont {A.~C.}\ \bibnamefont {Gossard}},\ }\href {\doibase
  10.1103/PhysRevLett.80.4522} {\bibfield  {journal} {\bibinfo  {journal}
  {Phys. Rev. Lett.}\ }\textbf {\bibinfo {volume} {80}},\ \bibinfo {pages}
  {4522} (\bibinfo {year} {1998})}\BibitemShut {NoStop}%
\bibitem [{\citenamefont {Duncan}\ \emph {et~al.}(2000)\citenamefont {Duncan},
  \citenamefont {Goldhaber-Gordon}, \citenamefont {Westervelt}, \citenamefont
  {Maranowski},\ and\ \citenamefont {Gossard}}]{duncan_2000}%
  \BibitemOpen
  \bibfield  {author} {\bibinfo {author} {\bibfnamefont {D.}~\bibnamefont
  {Duncan}}, \bibinfo {author} {\bibfnamefont {D.}~\bibnamefont
  {Goldhaber-Gordon}}, \bibinfo {author} {\bibfnamefont {R.}~\bibnamefont
  {Westervelt}}, \bibinfo {author} {\bibfnamefont {K.}~\bibnamefont
  {Maranowski}}, \ and\ \bibinfo {author} {\bibfnamefont {A.}~\bibnamefont
  {Gossard}},\ }\href@noop {} {\bibfield  {journal} {\bibinfo  {journal} {Appl.
  Phys. Lett.}\ }\textbf {\bibinfo {volume} {77}},\ \bibinfo {pages} {2183}
  (\bibinfo {year} {2000})}\BibitemShut {NoStop}%
\bibitem [{\citenamefont {Jezouin}\ \emph {et~al.}(2016)\citenamefont
  {Jezouin}, \citenamefont {Iftikhar}, \citenamefont {Anthore}, \citenamefont
  {Parmentier}, \citenamefont {Gennser}, \citenamefont {Cavanna}, \citenamefont
  {Ouerghi}, \citenamefont {Levkivskyi}, \citenamefont {Idrisov}, \citenamefont
  {Sukhorukov}, \citenamefont {Glazman},\ and\ \citenamefont
  {Pierre}}]{jezouin_2016}%
  \BibitemOpen
  \bibfield  {author} {\bibinfo {author} {\bibfnamefont {S.}~\bibnamefont
  {Jezouin}}, \bibinfo {author} {\bibfnamefont {Z.}~\bibnamefont {Iftikhar}},
  \bibinfo {author} {\bibfnamefont {A.}~\bibnamefont {Anthore}}, \bibinfo
  {author} {\bibfnamefont {F.~D.}\ \bibnamefont {Parmentier}}, \bibinfo
  {author} {\bibfnamefont {U.}~\bibnamefont {Gennser}}, \bibinfo {author}
  {\bibfnamefont {A.}~\bibnamefont {Cavanna}}, \bibinfo {author} {\bibfnamefont
  {A.}~\bibnamefont {Ouerghi}}, \bibinfo {author} {\bibfnamefont {I.~P.}\
  \bibnamefont {Levkivskyi}}, \bibinfo {author} {\bibfnamefont
  {E.}~\bibnamefont {Idrisov}}, \bibinfo {author} {\bibfnamefont {E.~V.}\
  \bibnamefont {Sukhorukov}}, \bibinfo {author} {\bibfnamefont {L.~I.}\
  \bibnamefont {Glazman}}, \ and\ \bibinfo {author} {\bibfnamefont
  {F.}~\bibnamefont {Pierre}},\ }\href@noop {} {\bibfield  {journal} {\bibinfo
  {journal} {Nature}\ }\textbf {\bibinfo {volume} {536}},\ \bibinfo {pages}
  {58} (\bibinfo {year} {2016})}\BibitemShut {NoStop}%
\bibitem [{\citenamefont {Averin}(1999)}]{Averin1999}%
  \BibitemOpen
  \bibfield  {author} {\bibinfo {author} {\bibfnamefont {D.~V.}\ \bibnamefont
  {Averin}},\ }\href {\doibase 10.1103/PhysRevLett.82.3685} {\bibfield
  {journal} {\bibinfo  {journal} {Phys. Rev. Lett.}\ }\textbf {\bibinfo
  {volume} {82}},\ \bibinfo {pages} {3685} (\bibinfo {year}
  {1999})}\BibitemShut {NoStop}%
\bibitem [{\citenamefont {Glazman}\ and\ \citenamefont
  {Matveev}(1989)}]{GlazmanMatveev1989}%
  \BibitemOpen
  \bibfield  {author} {\bibinfo {author} {\bibfnamefont {L.}~\bibnamefont
  {Glazman}}\ and\ \bibinfo {author} {\bibfnamefont {K.}~\bibnamefont
  {Matveev}},\ }\href {http://www.jetpletters.ac.ru/ps/1121/article_16988.pdf}
  {\bibfield  {journal} {\bibinfo  {journal} {JETP Lett.}\ }\textbf {\bibinfo
  {volume} {49}},\ \bibinfo {pages} {659} (\bibinfo {year} {1989})}\BibitemShut
  {NoStop}%
\bibitem [{\citenamefont {Beenakker}\ and\ \citenamefont {van
  Houten}(1992)}]{beenakkervanhouten}%
  \BibitemOpen
  \bibfield  {author} {\bibinfo {author} {\bibfnamefont {C.~W.~J.}\
  \bibnamefont {Beenakker}}\ and\ \bibinfo {author} {\bibfnamefont
  {H.}~\bibnamefont {van Houten}},\ }in\ \href
  {https://link.springer.com/chapter/10.1007/978-3-642-77274-0_20} {\emph
  {\bibinfo {booktitle} {Single-Electron Tunneling and Mesoscopic Devices}}},\
  \bibinfo {editor} {edited by\ \bibinfo {editor} {\bibfnamefont
  {H.}~\bibnamefont {Koch}}\ and\ \bibinfo {editor} {\bibfnamefont
  {H.}~\bibnamefont {L{\"u}bbig}}}\ (\bibinfo  {publisher} {Springer Berlin
  Heidelberg},\ \bibinfo {address} {Berlin, Heidelberg},\ \bibinfo {year}
  {1992})\ pp.\ \bibinfo {pages} {175--179}\BibitemShut {NoStop}%
\bibitem [{\citenamefont {Devyatov}\ and\ \citenamefont
  {Kupriyanov}(1997)}]{Devyatov1997}%
  \BibitemOpen
  \bibfield  {author} {\bibinfo {author} {\bibfnamefont {I.~A.}\ \bibnamefont
  {Devyatov}}\ and\ \bibinfo {author} {\bibfnamefont {M.~Y.}\ \bibnamefont
  {Kupriyanov}},\ }\href {\doibase 10.1134/1.558305} {\bibfield  {journal}
  {\bibinfo  {journal} {JETP}\ }\textbf {\bibinfo {volume} {85}},\ \bibinfo
  {pages} {189} (\bibinfo {year} {1997})}\BibitemShut {NoStop}%
\bibitem [{\citenamefont {Krogstrup}\ \emph {et~al.}(2015)\citenamefont
  {Krogstrup}, \citenamefont {Ziino}, \citenamefont {Chang}, \citenamefont
  {Albrecht}, \citenamefont {Madsen}, \citenamefont {Johnson}, \citenamefont
  {Nyg{\aa}rd}, \citenamefont {Marcus},\ and\ \citenamefont
  {Jespersen}}]{krogstrup_2015}%
  \BibitemOpen
  \bibfield  {author} {\bibinfo {author} {\bibfnamefont {P.}~\bibnamefont
  {Krogstrup}}, \bibinfo {author} {\bibfnamefont {N.~L.~B.}\ \bibnamefont
  {Ziino}}, \bibinfo {author} {\bibfnamefont {W.}~\bibnamefont {Chang}},
  \bibinfo {author} {\bibfnamefont {S.~M.}\ \bibnamefont {Albrecht}}, \bibinfo
  {author} {\bibfnamefont {M.~H.}\ \bibnamefont {Madsen}}, \bibinfo {author}
  {\bibfnamefont {E.}~\bibnamefont {Johnson}}, \bibinfo {author} {\bibfnamefont
  {J.}~\bibnamefont {Nyg{\aa}rd}}, \bibinfo {author} {\bibfnamefont {C.~M.}\
  \bibnamefont {Marcus}}, \ and\ \bibinfo {author} {\bibfnamefont {T.~S.}\
  \bibnamefont {Jespersen}},\ }\href@noop {} {\bibfield  {journal} {\bibinfo
  {journal} {Nat. Mater.}\ }\textbf {\bibinfo {volume} {14}},\ \bibinfo {pages}
  {400} (\bibinfo {year} {2015})}\BibitemShut {NoStop}%
\bibitem [{\citenamefont {Kringh{\o}j}\ \emph {et~al.}(2019)\citenamefont
  {Kringh{\o}j}, \citenamefont {Larsen}, \citenamefont {van Heck},
  \citenamefont {Sabonis}, \citenamefont {Erlandsson}, \citenamefont
  {Petkovic}, \citenamefont {Pikulin}, \citenamefont {Krogstrup}, \citenamefont
  {Petersson},\ and\ \citenamefont {Marcus}}]{lead}%
  \BibitemOpen
  \bibfield  {author} {\bibinfo {author} {\bibfnamefont {A.}~\bibnamefont
  {Kringh{\o}j}}, \bibinfo {author} {\bibfnamefont {T.~W.}\ \bibnamefont
  {Larsen}}, \bibinfo {author} {\bibfnamefont {B.}~\bibnamefont {van Heck}},
  \bibinfo {author} {\bibfnamefont {D.}~\bibnamefont {Sabonis}}, \bibinfo
  {author} {\bibfnamefont {O.}~\bibnamefont {Erlandsson}}, \bibinfo {author}
  {\bibfnamefont {I.}~\bibnamefont {Petkovic}}, \bibinfo {author}
  {\bibfnamefont {D.~I.}\ \bibnamefont {Pikulin}}, \bibinfo {author}
  {\bibfnamefont {P.}~\bibnamefont {Krogstrup}}, \bibinfo {author}
  {\bibfnamefont {K.~D.}\ \bibnamefont {Petersson}}, \ and\ \bibinfo {author}
  {\bibfnamefont {C.~M.}\ \bibnamefont {Marcus}},\ }\href@noop {} {\bibfield
  {journal} {\bibinfo  {journal} {arXiv:1910.08200}\ } (\bibinfo
  {year} {2019})}\BibitemShut {NoStop}%
\bibitem [{\citenamefont {Chang}\ \emph {et~al.}(2015)\citenamefont {Chang},
  \citenamefont {Albrecht}, \citenamefont {Jespersen}, \citenamefont
  {Kuemmeth}, \citenamefont {Krogstrup}, \citenamefont {Nyg{\aa}rd},\ and\
  \citenamefont {Marcus}}]{chang_2015}%
  \BibitemOpen
  \bibfield  {author} {\bibinfo {author} {\bibfnamefont {W.}~\bibnamefont
  {Chang}}, \bibinfo {author} {\bibfnamefont {S.~M.}\ \bibnamefont {Albrecht}},
  \bibinfo {author} {\bibfnamefont {T.~S.}\ \bibnamefont {Jespersen}}, \bibinfo
  {author} {\bibfnamefont {F.}~\bibnamefont {Kuemmeth}}, \bibinfo {author}
  {\bibfnamefont {P.}~\bibnamefont {Krogstrup}}, \bibinfo {author}
  {\bibfnamefont {J.}~\bibnamefont {Nyg{\aa}rd}}, \ and\ \bibinfo {author}
  {\bibfnamefont {C.~M.}\ \bibnamefont {Marcus}},\ }\href@noop {} {\bibfield
  {journal} {\bibinfo  {journal} {Nat. Nanotechnol.}\ }\textbf {\bibinfo
  {volume} {10}},\ \bibinfo {pages} {232} (\bibinfo {year} {2015})}\BibitemShut
  {NoStop}%
\bibitem [{\citenamefont {Hart}\ \emph {et~al.}(2019)\citenamefont {Hart},
  \citenamefont {Cui}, \citenamefont {M\'enard}, \citenamefont {Deng},
  \citenamefont {Antipov}, \citenamefont {Lutchyn}, \citenamefont {Krogstrup},
  \citenamefont {Marcus},\ and\ \citenamefont {Moler}}]{Hart2019}%
  \BibitemOpen
  \bibfield  {author} {\bibinfo {author} {\bibfnamefont {S.}~\bibnamefont
  {Hart}}, \bibinfo {author} {\bibfnamefont {Z.}~\bibnamefont {Cui}}, \bibinfo
  {author} {\bibfnamefont {G.}~\bibnamefont {M\'enard}}, \bibinfo {author}
  {\bibfnamefont {M.}~\bibnamefont {Deng}}, \bibinfo {author} {\bibfnamefont
  {A.~E.}\ \bibnamefont {Antipov}}, \bibinfo {author} {\bibfnamefont {R.~M.}\
  \bibnamefont {Lutchyn}}, \bibinfo {author} {\bibfnamefont {P.}~\bibnamefont
  {Krogstrup}}, \bibinfo {author} {\bibfnamefont {C.~M.}\ \bibnamefont
  {Marcus}}, \ and\ \bibinfo {author} {\bibfnamefont {K.~A.}\ \bibnamefont
  {Moler}},\ }\href {\doibase 10.1103/PhysRevB.100.064523} {\bibfield
  {journal} {\bibinfo  {journal} {Phys. Rev. B}\ }\textbf {\bibinfo {volume}
  {100}},\ \bibinfo {pages} {064523} (\bibinfo {year} {2019})}\BibitemShut
  {NoStop}%
\bibitem [{sup()}]{supplement}%
  \BibitemOpen
  \href@noop {} {}\bibinfo {note} {See Supplemental Material for details on the
  theoretical modelling, transport measurements, data extraction, and open
  regime charge dispersion.}\BibitemShut {Stop}%
\bibitem [{\citenamefont {Larkin}\ and\ \citenamefont
  {Matveev}(1987)}]{LarkinMatveev1987}%
  \BibitemOpen
  \bibfield  {author} {\bibinfo {author} {\bibfnamefont {A.}~\bibnamefont
  {Larkin}}\ and\ \bibinfo {author} {\bibfnamefont {K.}~\bibnamefont
  {Matveev}},\ }\href {http://www.jetp.ac.ru/cgi-bin/dn/e_066_03_0580.pdf}
  {\bibfield  {journal} {\bibinfo  {journal} {JETP}\ }\textbf {\bibinfo
  {volume} {66}},\ \bibinfo {pages} {590} (\bibinfo {year} {1987})}\BibitemShut
  {NoStop}%
\bibitem [{\citenamefont {Ivanov}\ and\ \citenamefont
  {Feigel'man}(1998)}]{Ivanov1998}%
  \BibitemOpen
  \bibfield  {author} {\bibinfo {author} {\bibfnamefont {D.~A.}\ \bibnamefont
  {Ivanov}}\ and\ \bibinfo {author} {\bibfnamefont {M.~V.}\ \bibnamefont
  {Feigel'man}},\ }\href {\doibase 10.1134/1.558666} {\bibfield  {journal}
  {\bibinfo  {journal} {JETP}\ }\textbf {\bibinfo {volume} {87}},\ \bibinfo
  {pages} {349} (\bibinfo {year} {1998})}\BibitemShut {NoStop}%
\bibitem [{\citenamefont {Ivanov}\ and\ \citenamefont
  {Feigel'man}(1999)}]{Ivanov1999}%
  \BibitemOpen
  \bibfield  {author} {\bibinfo {author} {\bibfnamefont {D.~A.}\ \bibnamefont
  {Ivanov}}\ and\ \bibinfo {author} {\bibfnamefont {M.~V.}\ \bibnamefont
  {Feigel'man}},\ }\href {\doibase 10.1103/PhysRevB.59.8444} {\bibfield
  {journal} {\bibinfo  {journal} {Phys. Rev. B}\ }\textbf {\bibinfo {volume}
  {59}},\ \bibinfo {pages} {8444} (\bibinfo {year} {1999})}\BibitemShut
  {NoStop}%
\bibitem [{\citenamefont {Zazunov}\ \emph {et~al.}(2005)\citenamefont
  {Zazunov}, \citenamefont {Shumeiko}, \citenamefont {Wendin},\ and\
  \citenamefont {Bratus'}}]{Zazunov2005}%
  \BibitemOpen
  \bibfield  {author} {\bibinfo {author} {\bibfnamefont {A.}~\bibnamefont
  {Zazunov}}, \bibinfo {author} {\bibfnamefont {V.~S.}\ \bibnamefont
  {Shumeiko}}, \bibinfo {author} {\bibfnamefont {G.}~\bibnamefont {Wendin}}, \
  and\ \bibinfo {author} {\bibfnamefont {E.~N.}\ \bibnamefont {Bratus'}},\
  }\href {\doibase 10.1103/PhysRevB.71.214505} {\bibfield  {journal} {\bibinfo
  {journal} {Phys. Rev. B}\ }\textbf {\bibinfo {volume} {71}},\ \bibinfo
  {pages} {214505} (\bibinfo {year} {2005})}\BibitemShut {NoStop}%
\bibitem [{Note1()}]{Note1}%
  \BibitemOpen
  \bibinfo {note} {Due to the large area below the potential barrier for $4\pi
  $-tunneling at perfect transmission, this residual dispersion can be
  estimated to be well below the experimentally achieved
  linewidth.}\BibitemShut {Stop}%
\bibitem [{com()}]{comsol}%
  \BibitemOpen
  \href@noop {} {}\bibinfo {note} {COMSOL, Inc. [www.comsol.com]}\BibitemShut
  {NoStop}%
\bibitem [{\citenamefont {Doh}\ \emph {et~al.}(2005)\citenamefont {Doh},
  \citenamefont {van Dam}, \citenamefont {Roest}, \citenamefont {Bakkers},
  \citenamefont {Kouwenhoven},\ and\ \citenamefont {De~Franceschi}}]{doh_2005}%
  \BibitemOpen
  \bibfield  {author} {\bibinfo {author} {\bibfnamefont {Y.-J.}\ \bibnamefont
  {Doh}}, \bibinfo {author} {\bibfnamefont {J.~A.}\ \bibnamefont {van Dam}},
  \bibinfo {author} {\bibfnamefont {A.~L.}\ \bibnamefont {Roest}}, \bibinfo
  {author} {\bibfnamefont {E.~P. A.~M.}\ \bibnamefont {Bakkers}}, \bibinfo
  {author} {\bibfnamefont {L.~P.}\ \bibnamefont {Kouwenhoven}}, \ and\ \bibinfo
  {author} {\bibfnamefont {S.}~\bibnamefont {De~Franceschi}},\ }\href {\doibase
  10.1126/science.1113523} {\bibfield  {journal} {\bibinfo  {journal}
  {Science}\ }\textbf {\bibinfo {volume} {309}},\ \bibinfo {pages} {272}
  (\bibinfo {year} {2005})}\BibitemShut {NoStop}%
\bibitem [{Note2()}]{Note2}%
  \BibitemOpen
  \bibinfo {note} {Numerical code and data accompanying the analysis of Figs.~4
  and 5 are found at:\\ \protect \url
  {https://github.com/anderskringhoej/Dispersion}.}\BibitemShut {Stop}%
\bibitem [{\citenamefont {Vecino}\ \emph {et~al.}(2003)\citenamefont {Vecino},
  \citenamefont {Mart\'{\i}n-Rodero},\ and\ \citenamefont
  {Levy~Yeyati}}]{Vecino2003}%
  \BibitemOpen
  \bibfield  {author} {\bibinfo {author} {\bibfnamefont {E.}~\bibnamefont
  {Vecino}}, \bibinfo {author} {\bibfnamefont {A.}~\bibnamefont
  {Mart\'{\i}n-Rodero}}, \ and\ \bibinfo {author} {\bibfnamefont
  {A.}~\bibnamefont {Levy~Yeyati}},\ }\href {\doibase
  10.1103/PhysRevB.68.035105} {\bibfield  {journal} {\bibinfo  {journal} {Phys.
  Rev. B}\ }\textbf {\bibinfo {volume} {68}},\ \bibinfo {pages} {035105}
  (\bibinfo {year} {2003})}\BibitemShut {NoStop}%
\bibitem [{\citenamefont {Mart{\'\i}n-Rodero}\ and\ \citenamefont
  {Levy~Yeyati}(2011)}]{MartinRodero2011}%
  \BibitemOpen
  \bibfield  {author} {\bibinfo {author} {\bibfnamefont {A.}~\bibnamefont
  {Mart{\'\i}n-Rodero}}\ and\ \bibinfo {author} {\bibfnamefont
  {A.}~\bibnamefont {Levy~Yeyati}},\ }\href
  {https://www.tandfonline.com/doi/abs/10.1080/00018732.2011.624266} {\bibfield
   {journal} {\bibinfo  {journal} {Advances in Physics}\ }\textbf {\bibinfo
  {volume} {60}},\ \bibinfo {pages} {899} (\bibinfo {year} {2011})}\BibitemShut
  {NoStop}%
\bibitem [{\citenamefont {Janvier}\ \emph {et~al.}(2015)\citenamefont
  {Janvier}, \citenamefont {Tosi}, \citenamefont {Bretheau}, \citenamefont
  {Girit}, \citenamefont {Stern}, \citenamefont {Bertet}, \citenamefont
  {Joyez}, \citenamefont {Vion}, \citenamefont {Esteve}, \citenamefont
  {Goffman}, \citenamefont {Pothier},\ and\ \citenamefont
  {Urbina}}]{janvier_2016}%
  \BibitemOpen
  \bibfield  {author} {\bibinfo {author} {\bibfnamefont {C.}~\bibnamefont
  {Janvier}}, \bibinfo {author} {\bibfnamefont {L.}~\bibnamefont {Tosi}},
  \bibinfo {author} {\bibfnamefont {L.}~\bibnamefont {Bretheau}}, \bibinfo
  {author} {\bibfnamefont {{\c C}.~{\"O}.}\ \bibnamefont {Girit}}, \bibinfo
  {author} {\bibfnamefont {M.}~\bibnamefont {Stern}}, \bibinfo {author}
  {\bibfnamefont {P.}~\bibnamefont {Bertet}}, \bibinfo {author} {\bibfnamefont
  {P.}~\bibnamefont {Joyez}}, \bibinfo {author} {\bibfnamefont
  {D.}~\bibnamefont {Vion}}, \bibinfo {author} {\bibfnamefont {D.}~\bibnamefont
  {Esteve}}, \bibinfo {author} {\bibfnamefont {M.~F.}\ \bibnamefont {Goffman}},
  \bibinfo {author} {\bibfnamefont {H.}~\bibnamefont {Pothier}}, \ and\
  \bibinfo {author} {\bibfnamefont {C.}~\bibnamefont {Urbina}},\ }\href
  {\doibase 10.1126/science.aab2179} {\bibfield  {journal} {\bibinfo  {journal}
  {Science}\ }\textbf {\bibinfo {volume} {349}},\ \bibinfo {pages} {1199}
  (\bibinfo {year} {2015})}\BibitemShut {NoStop}%
\bibitem [{\citenamefont {Hays}\ \emph {et~al.}(2018)\citenamefont {Hays},
  \citenamefont {de~Lange}, \citenamefont {Serniak}, \citenamefont {van
  Woerkom}, \citenamefont {Bouman}, \citenamefont {Krogstrup}, \citenamefont
  {Nyg\aa{}rd}, \citenamefont {Geresdi},\ and\ \citenamefont
  {Devoret}}]{Hays_2018}%
  \BibitemOpen
  \bibfield  {author} {\bibinfo {author} {\bibfnamefont {M.}~\bibnamefont
  {Hays}}, \bibinfo {author} {\bibfnamefont {G.}~\bibnamefont {de~Lange}},
  \bibinfo {author} {\bibfnamefont {K.}~\bibnamefont {Serniak}}, \bibinfo
  {author} {\bibfnamefont {D.~J.}\ \bibnamefont {van Woerkom}}, \bibinfo
  {author} {\bibfnamefont {D.}~\bibnamefont {Bouman}}, \bibinfo {author}
  {\bibfnamefont {P.}~\bibnamefont {Krogstrup}}, \bibinfo {author}
  {\bibfnamefont {J.}~\bibnamefont {Nyg\aa{}rd}}, \bibinfo {author}
  {\bibfnamefont {A.}~\bibnamefont {Geresdi}}, \ and\ \bibinfo {author}
  {\bibfnamefont {M.~H.}\ \bibnamefont {Devoret}},\ }\href {\doibase
  10.1103/PhysRevLett.121.047001} {\bibfield  {journal} {\bibinfo  {journal}
  {Phys. Rev. Lett.}\ }\textbf {\bibinfo {volume} {121}},\ \bibinfo {pages}
  {047001} (\bibinfo {year} {2018})}\BibitemShut {NoStop}%
\bibitem [{\citenamefont {Bargerbos}\ \emph {et~al.}(2019)\citenamefont
  {Bargerbos}, \citenamefont {Uilhoorn}, \citenamefont {Yang}, \citenamefont
  {Krogstrup}, \citenamefont {Kouwenhoven}, \citenamefont {de~Lange},
  \citenamefont {van Heck},\ and\ \citenamefont {Kou}}]{arno}%
  \BibitemOpen
  \bibfield  {author} {\bibinfo {author} {\bibfnamefont {A.}~\bibnamefont
  {Bargerbos}}, \bibinfo {author} {\bibfnamefont {W.}~\bibnamefont {Uilhoorn}},
  \bibinfo {author} {\bibfnamefont {C.-K.}\ \bibnamefont {Yang}}, \bibinfo
  {author} {\bibfnamefont {P.}~\bibnamefont {Krogstrup}}, \bibinfo {author}
  {\bibfnamefont {L.~P.}\ \bibnamefont {Kouwenhoven}}, \bibinfo {author}
  {\bibfnamefont {G.}~\bibnamefont {de~Lange}}, \bibinfo {author}
  {\bibfnamefont {B.}~\bibnamefont {van Heck}}, \ and\ \bibinfo {author}
  {\bibfnamefont {A.}~\bibnamefont {Kou}},\ }\href@noop {} {\bibfield
  {journal} {\bibinfo  {journal} {arXiv:1911.10010}\ } (\bibinfo
  {year} {2019})}\BibitemShut {NoStop}%
\end{thebibliography}
\enlargethispage{3\baselineskip}
\let\oldaddcontentsline\addcontentsline% Store \addcontentsline
\renewcommand{\addcontentsline}[3]{}% Make \addcontentsline a no-op
\let\addcontentsline\oldaddcontentsline% Restore \addcontentsline

\newcommand{\beginsupplement}{%
        \setcounter{table}{0}
        \renewcommand{\thetable}{S\arabic{table}}%
        \setcounter{figure}{0}
        \renewcommand{\thefigure}{S\arabic{figure}}%
     }

\onecolumngrid
\let\oldaddcontentsline\addcontentsline% Store \addcontentsline
\renewcommand{\addcontentsline}[3]{}% Make \addcontentsline a no-op
\section{Supplementary Material}
\let\addcontentsline\oldaddcontentsline% Restore \addcontentsline
%\maketitle

\beginsupplement
\setcounter{tocdepth}{2}
\tableofcontents

\section{Theory}

\subsection{Derivation of the bound state equation}

The bound state equation, Eq.~(1) in the main text, has been previously derived within a scattering matrix formalism~\cite{beenakkervanhouten}.
For completeness, we present here an alternative derivation based on the tunneling Hamiltonian.
Namely, we consider the following model of a Josephson junction with a resonant level coupling two $s$-wave superconductors,
\begin{subequations}
\begin{align}
H&=H_0 + H_\textrm{tunn}\;,\\
H_0&=\epsilon_r\, \sum_\sigma d^\dagger_{\sigma} d_\sigma + \sum_{\alpha n \sigma} E_{\alpha n} \Gamma^\dagger_{\alpha n \sigma}\Gamma_{\alpha n \sigma}\;,\\
H_\textrm{tunn} & = \sum_\alpha \e^{-i\phi_\alpha/2}\,t_\alpha\,\sum_{n \sigma}\,\left[u_{\alpha n}\, d^\dagger_\sigma\Gamma_{\alpha n \sigma} + \sigma v_{\alpha n} \,d^\dagger_\sigma \Gamma^\dagger_{\alpha n \bar\sigma}\right]\,+\,\textrm{h.c.}
\end{align}
\end{subequations}
Here, $H_0$ is the Hamiltonian in the absence of tunneling between the dot and the leads and $H_\textrm{tunn}$ is the tunneling Hamiltonian; $\epsilon_r$ is the energy of the resonant level; $\alpha=1,2$ labels the two leads; $n$ labels the orbitals in the two leads; $\sigma$ labels spin; $\bar{\sigma}=-\sigma$; $\phi_\alpha$ is the superconducting phase in lead $\alpha$; $t_\alpha$ is the tunneling strength between the dot and the lead $\alpha$; $u_{\alpha n}$ and $v_{\alpha n}$ are the BCS coherence factors for the quasiparticle states in the leads.
We have assumed for simplicity that the tunneling strength is identical for every quasiparticle state in each lead, and that spin is a good quantum number.

The single-particle excitation energies of the Hamiltonian $H$ are the positive energy solutions of the Bogoliubov-de Gennes equations $H_\textrm{BdG}\Psi=E\Psi$, derived by rewriting the Hamiltonian in Nambu (particle/hole) space.
Here, $\Psi=(\Phi, \tilde{\Phi})$ is a Nambu wave function, and both $\Phi$ and $\tilde{\Phi}$ have components on the resonant level (which we will denote by $\Phi_0, \tilde{\Phi_0}$) as well as on the quasiparticle levels (which we will denote by $\Phi_{\alpha n}, \tilde{\Phi}_{\alpha n}$) 
The Bogoliubov-de Gennes equations are explicitly given by
\begin{subequations}
\begin{align}
\sum_{\alpha n} u_{\alpha n}t_\alpha \e^{-i\phi_\alpha/2} \Phi_{\alpha n} + \sum_{\alpha n} v_{\alpha n} t_\alpha \e^{-i\phi_\alpha/2} \tilde{\Phi}_{\alpha n} &= (E-\epsilon_r)\,\Phi_0\\
u_{\alpha n}t_\alpha \e^{i\phi_\alpha/2} \Phi_0+ v_{\alpha n} t_\alpha \e^{-i\phi_\alpha/2} \tilde{\Phi}_0 & = (E-\epsilon_r)\,\Phi_{\alpha n}\\
- \sum_{\alpha n} u_{\alpha n}t_\alpha \e^{i\phi_\alpha/2} \tilde{\Phi}_{\alpha n} + \sum_{\alpha n} v_{\alpha n} t_\alpha \e^{i\phi_\alpha/2} \Phi_{\alpha n} &= (E+\epsilon_r) \,\tilde{\Phi}_0\\
- u_{\alpha n}t_\alpha \e^{-i\phi_\alpha/2} \tilde\Phi_0+ v_{\alpha n} t_\alpha \e^{i\phi_\alpha/2} \Phi_0 & = (E+\epsilon_r)\,\tilde{\Phi}_{\alpha n}.
\end{align}
\end{subequations}
Note that the spin indices $\sigma$ have been suppressed since they play a trivial role because spin is conserved by $H$. From Eq.~(2b) and Eq.~(2d), we can express the quasiparticle components in terms of $\Phi_0, \tilde{\Phi}_0$,
\begin{subequations}
\begin{align}
\Phi_{\alpha n}&=\frac{u_{\alpha n}t_\alpha \e^{i\phi_\alpha/2}}{E-E_{\alpha n}} \Phi_0+ \frac{v_{\alpha n} t_\alpha \e^{-i\phi_\alpha/2}}{E-E_{\alpha n}} \tilde{\Phi}_0\,\\
\tilde{\Phi}_{\alpha n}&=\frac{v_{\alpha n}t_\alpha \e^{i\phi_\alpha/2}}{E+E_{\alpha n}} \Phi_0- \frac{u_{\alpha n} t_\alpha \e^{-i\phi_\alpha/2}}{E+E_{\alpha n}} \tilde{\Phi}_0\,.
\end{align}
\end{subequations}
We can now insert Eq.~(3a) and Eq.~(3b) into Eq.~(2a) and Eq.~(2c), which results in a $2\times 2$ system of linear equations that only involves $\Phi_0$ and $\tilde\Phi_0$.
\begin{subequations}\label{eq:2x2_bound_state_eqs}
\begin{align}
A(E)\,\Phi_0 + B(E)\,\tilde{\Phi}_0 &= (E-\epsilon_r)\,\Phi_0\\
B^*(E)\,\Phi_0 + A(E)\,\tilde{\Phi}_0 &= (E+\epsilon_r)\,\tilde{\Phi}_0.
\end{align}
\end{subequations}
The coefficients are energy-dependent:
\begin{subequations}
\begin{align}
A(E)&=-\sum_{\alpha} \Gamma_\alpha \,\frac{E}{\sqrt{\Delta^2-E^2}}\\
B(E)&=-\sum_{\alpha} \Gamma_\alpha\e^{-i\phi_\alpha}\,\frac{\Delta}{\sqrt{\Delta^2-E^2}}.
\end{align}
\end{subequations}
They can be derived using the expressions for $u_{\alpha n}$ and $v_{\alpha n}$, namely $u^2_n = \tfrac{1}{2}\left(1+\xi_n/\epsilon_n\right)$ and $v^2_n = \tfrac{1}{2}\left(1-\xi_n/\epsilon_n\right)$ with $\epsilon_n=\sqrt{\xi_n^2+\Delta^2}$ and by performing the sums over $n$ in the continuum limit (the resulting integrals converge for $E<\Delta$). In the above equation we have introduced the tunneling rates
\begin{equation}
\Gamma_\alpha = \frac{\pi t^2_\alpha}{\delta_\alpha}\,.
\end{equation}
The $2\times 2$ system of equations~\eqref{eq:2x2_bound_state_eqs} has a solution if
\begin{equation}
[A(E)-(E-\epsilon_r)][D(E)-(E+\epsilon_r)] - \abs{B(E)}^2 = 0.
\end{equation}
This amounts to the bound state equation quoted in the main text,
\begin{equation}\label{eq:bound_state_eq}
2\sqrt{\Delta^2-E^2}\,E^2\,\Gamma + (\Delta^2-E^2)(E^2-\epsilon_r^2-\Gamma^2) + 4\Delta^2 \Gamma_1\Gamma_2\,\sin^2(\phi/2) = 0,
\end{equation}
where $\Gamma=\Gamma_1+\Gamma_2$ and $\phi=\phi_2-\phi_1$.
This equation is equal to the one reported in Refs.~\cite{beenakkervanhouten,golubov_2004}, up to the fact that $\Gamma_\alpha$ are defined here without a factor of two associated with spin degeneracy.

\subsection{Properties of the bound state energy}

\begin{figure}
    \centering{}
        \includegraphics[width=1\textwidth]{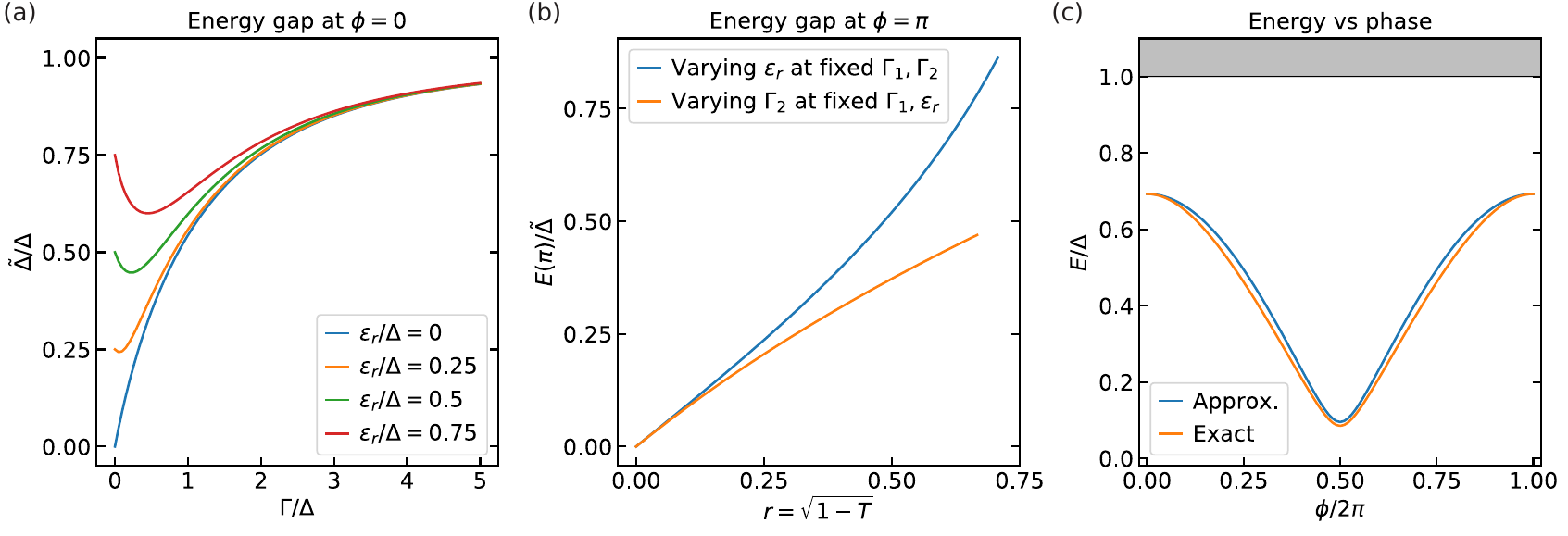}
    \caption{Properties of the bound state energy obtained from the resonant level model. (a) Energy gap at $\phi=0$ ($\tilde{\Delta}$) for different values of the coupling $\Gamma=\Gamma_1+\Gamma_2$ and the resonant level energy $\epsilon_r$. (b) Parametric plot of the energy gap at $\phi=\pi$ for increasing values of the reflection amplitude $r$. Both quantities were computed by either varying $\epsilon_r$ for a fixed symmetric coupling $\Gamma_1=\Gamma_2=\tfrac{1}{2}\Delta$, or by varying $\Gamma_2$ at fixed $\epsilon_r=0$, $\Gamma_1=\tfrac{1}{2}\Delta$. (c) Phase dependence of the bound state energy for the couplings used in the fit for Res.~1 of the main text: $\Delta/h=45$\,GHz and $\Gamma/2h=\Gamma_1/h=\Gamma_2/h=36$\,GHz. We fixed $\epsilon_r/h$ to a representative value of $10$\,GHz. The exact solution is a numerical solution of Eq.~\eqref{eq:bound_state_eq}. The approximate solution is given by $E=\tilde{\Delta}\sqrt{1-T\sin^2(\phi/2)}$.
    }
    \label{fig:bound_state_properties}
\end{figure}  

Here we discuss the properties of the solutions of Eq.~\eqref{eq:bound_state_eq}; see also Ref.~\cite{golubov_2004}.
Introducing the Breit-Wigner transmission through the resonant level at $E=0$, $T = 4\Gamma_1\Gamma_2/(\epsilon_r^2+\Gamma^2)$,
the bound state equation can be written in the convenient form
\begin{equation}\label{eq:bound_state_eq_revisited}
E^2\,[1+f(E)]=\Delta^2\,(1-T\sin^2\phi/2)
\end{equation}
with
\begin{equation}
f(E) =\frac{2\Gamma\sqrt{\Delta^2-E^2}}{\epsilon^2_r+\Gamma^2}+\frac{\Delta^2-E^2}{\epsilon^2_r+\Gamma^2},
\end{equation}
a dimensionless positive function, defined in the interval $0\leq E < \Delta$, which decreases monotonously from a value $f(0) = (\Delta^2 + 2\Gamma\Delta)/(\epsilon^2_r + \Gamma^2)$ to $f(\Delta)=0$.

In the main text we have defined the bound state energy at zero phase difference as $\tilde{\Delta}$.
It can be seen easily that $\tilde{\Delta}$ only depends on the total coupling $\Gamma$ and is thus insensitive to coupling asymmetry.
It is determined by the equation $\tilde{\Delta}^2 [1+f(\tilde{\Delta})]=\Delta^2$, which has approximate solutions $\tilde{\Delta}\approx\Gamma$ for $\Gamma\ll\Delta$ and $\tilde\Delta\approx\Delta$ for $\Gamma\gg\Delta$.
The complete behavior of $\tilde{\Delta}$ as a function of $\Gamma$ and $\epsilon_r$, obtained from a numerical solution of the bound state equation, is illustrated in Fig.~\ref{fig:bound_state_properties}(a).

The minimum bound state energy is always achieved at $\phi=\pi$, when the right hand side of Eq.~\eqref{eq:bound_state_eq_revisited} is minimized.
In particular, Eq.~\eqref{eq:bound_state_eq_revisited} shows that $E(\pi)=0$ for $T=1$ independent of the value of other parameters, and that $E(\pi)\approx \tilde{\Delta} \sqrt{1-T}$ for $\sqrt{1-T}\ll 1$.
The behavior of $E(\pi)$ as a function of $T$ is shown Fig.~\ref{fig:bound_state_properties}(b), obtained by varying either the asymmetry between the couplings ($\Gamma_1-\Gamma_2$) or $\epsilon_r$.

As Eq.~\eqref{eq:bound_state_eq_revisited} suggests, the entire phase dependence of the bound state energy is very well approximated by $E=\tilde{\Delta}[1-T\sin^2(\phi/2)]$. This relation becomes exact in the two opposite limits $f(0)\ll 1$ (which happens for $\epsilon_r^2+\Gamma^2\gg \Delta$), where $\tilde\Delta\approx \Delta$, and $f(0)\gg 1$ (for $\epsilon_r^2+\Gamma^2\ll\Delta$), where $\tilde{\Delta}\approx \Gamma$. For intermediate values of $\Gamma$, the regime where the optimal fit to the experimental data lies, the agreement is still very good, as shown in Fig.~\ref{fig:bound_state_properties}(c). This justifies the use of the model of Eq.~(2) in the main text for the calculation of the qubit levels.

\subsection{Qubit energy levels: numerical solutions}

The Hamiltonian Eq. (2) in the main text is used to the determine the qubit energy levels given the input parameters $E_C,~n_g,~\tilde{\Delta}$ and $r=\sqrt{1-T}$.
The Hamiltonian is solved numerically by discretizing the coordinate $\phi$ on a finite grid with grid spacing $\delta$, chosen to be small enough to guarantee convergence of the eigenvalues.
Following standard procedure, the derivative operator $\partial_\phi$ is implemented as a hopping operator between neighboring sites of the $\phi$-grid, with hopping strength $4 E_C / \delta^2$.
The induced charge $n_g$ enters the Hamiltonian, via the Peierls substitution, as a hopping phase $\e^{i\delta n_g/2}$.
We diagonalize the Hamiltonian on the interval $\phi\,\in\,[0, 4\pi)$ with anti-periodic boundary conditions.
This choice is required to guarantee the smoothness of the wave functions and the correct offset of energy levels with respect to $n_g$.

\section{Transport measurements}

\begin{figure}[ht]
    \centering
        \includegraphics[width=1\textwidth]{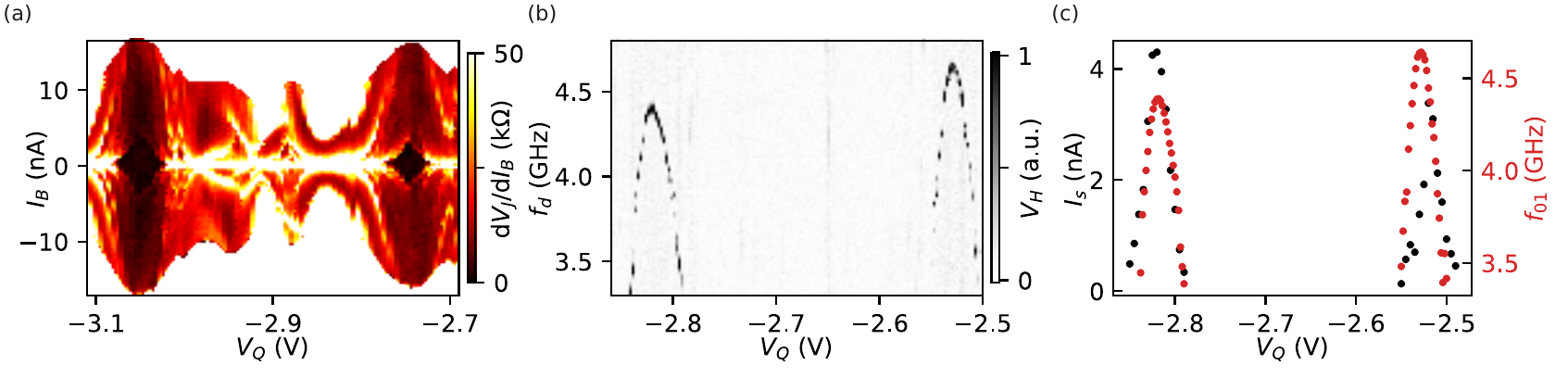}\vspace{-2mm}
    \caption{\vspace{-0mm} Comparison of DC transport and cQED measurements across the two resonances. (a) Measurement of the differential resistance d$V_J/$d$I_B$ as a function of current bias $I_B$ and $V_Q$ across the two resonances. The measurement is performed at $V_\text{FET}=+4$\,V where the FET is fully conducting such that the gate voltage across the nanowire effectively drops across the qubit junction. Two regions of supercurrent are observed. We identify the $I_B$ where the junction change from the non-resistive state to the resistive state as the switching current $I_s$. (b) Two-tone spectroscopy data across both resonances measured at $V_\text{FET}=-4$\,V completely depleting the FET. Qubit frequencies $f_{01}$ are extracted by Lorentzian fits. (c) Comparison of the extracted $I_s$ from (a) plotted on the left y-axis and the extracted $f_{01}$ from (b). Due to gate drift the $I_s$ curve is shifted by 200\,mV to align the resonance peaks.
    }
    \label{fig:supp transport}
\end{figure}

As the device described in the main text also has the capability of measuring transport when the FET is opened~\cite{lead}, we studied the resonances in DC transport measurements. At $V_\text{FET}=+4\,$V, when the FET was fully conducting, we measured the current $I_B$ and d$I_B$/d$V_B$ as a function of voltage bias $V_B$ and $V_Q$ across the resonances. By inverting d$I_B/$d$V_B$ and subtracting the line resistance $R=57\,$k$\Omega$ we infer the differential resistance across the qubit junction d$V_J/$d$I_B$ as shown in Fig.~\ref{fig:supp transport}(a). Here $V_J$ is the voltage drop across the qubit junction. From this measurement the switching current $I_s$ is extracted. $I_s$ is defined as the maximal value of d$V_J/$d$I_B$ before the junction turns from being in the non-resistive to the resistive state. In Fig.~\ref{fig:supp transport}(b) we plot the two-tone spectroscopy measurement across the resonances, which is also presented in Fig.~1(c) in the main text.
This allows us to compare the extracted $I_s$ with the extracted $f_{01}$ across the resonances, Fig.~\ref{fig:supp transport}(c). Here we observe a resonance structure the measured $I_s$ of similar width and spacing as the $f_{01}$. This further supports the interpretation of resonant tunneling through a single dot level~\cite{beenakkervanhouten}. In this comparison, $V_Q$ is shifted $\sim200$\,mV for the measurements of $I_s$  to align the resonances. We attribute this to gate drift common to these devices and crosstalk between the two gates as the FET is being varied from conducting to non-conducting.

\section{Data extraction}

\begin{figure}
    \centering
        \includegraphics[width=1\textwidth]{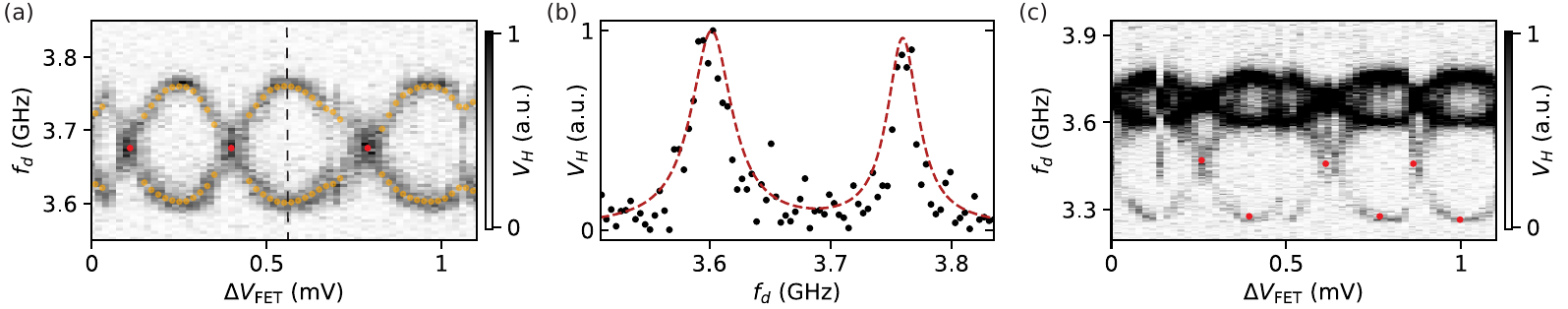}\vspace{-2mm}
    \caption{\vspace{-0mm} Examples of charge dispersion extraction. (a) An example of one of the datasets used to extract $\delta_{01}$ of Res.~2. By sweeping $V_\text{FET}$ in a small range we change the offset charge $n_g$ without varying $f_{01}$. $\Delta V_\text{FET}$ refer to the variation in $V_\text{FET}$ from its usual operation point $V_\text{FET}=-4$\,V. Varying $V_\text{FET}$ over such small voltages allows changing $n_g$ while keeping the FET depleted. By fitting each line to a double Lorentzians, we extract the two frequency branches (orange data points). Local maxima allow identifying $f_+$. The qubit degeneracy frequency $f_{01}$ (red data points) is extracted by fits to single Lorentzians. For Res.~2 an average of the extracted data points results in the extracted $f_{01}$ and $\delta_{01}$ for each $V_Q$. For Res.~1 each extracted value correspond to one data point as $f_{01}$ is varied together with $n_g$. (b) An example of a double Lorentzian fit used to identify the orange points in (a). The dashed line in (a) indicates $\Delta V_\text{FET}$ for the fitted dataset. (c) An example of a high power measurement of $\delta_{02}/2$ used to extract the data points in Fig.~5 of the main text. The lower frequency branch of the $0\to2$ is extracted by fits to single Lorentzians. The degeneracy frequency $f_{02}/2$ is manually estimated (red data points). As in (a) an average of the extracted values result in the extracted $f_{02}/2$ and $\delta_{02}/2$.
    }
    \label{fig:extraction}
\end{figure}

Dispersion data were measured by varying $n_g$, either with $V_Q$ (Res.~1) or $V_\text{FET}$ (Res.~2). An example of a dataset is shown in Fig.~\ref{fig:extraction}(a). Here the frequencies of the even and odd branches are extracted by fits to a double Lorentzians for each $\Delta V_\text{FET}$ (For Res.~1 frequencies are extracted with fits for each $V_Q$).  Here $\Delta V_\text{FET}$ refers to the voltage change in $V_\text{FET}$ away from the static operation point at $V_\text{FET}=-3$\,V, where the FET is fully depleted. An example of a fit is shown in Fig.~\ref{fig:extraction}(b).  We extract $f_+$ and $f_-$ at gate values of local maxima of their difference. We extract the degeneracy qubit frequency $f_{01}$ by fits to a single Lorentzian at $\Delta V_\text{FET}$ where the odd and even branches cross.

For Res.~2 the $0\to2$ two-photon transitions frequencies are also extracted. An example is shown in Fig.~\ref{fig:extraction}(c) where it is evident that the lower branch of the $0\to1$ interferes with the upper branch of the $0\to2$ transition. However, as both the degeneracy and minimal frequency are clearly distinguishable we define $\delta_{02}=f_{02}-f_{02,-}$. $f_{02,-}$ is extracted by fits to a single Lorentzian and $f_{02}$ is extracted manually.

\section{Spectroscopy and charge dispersion in the open regime}

\begin{figure}
    \centering
        \includegraphics[width=1\textwidth]{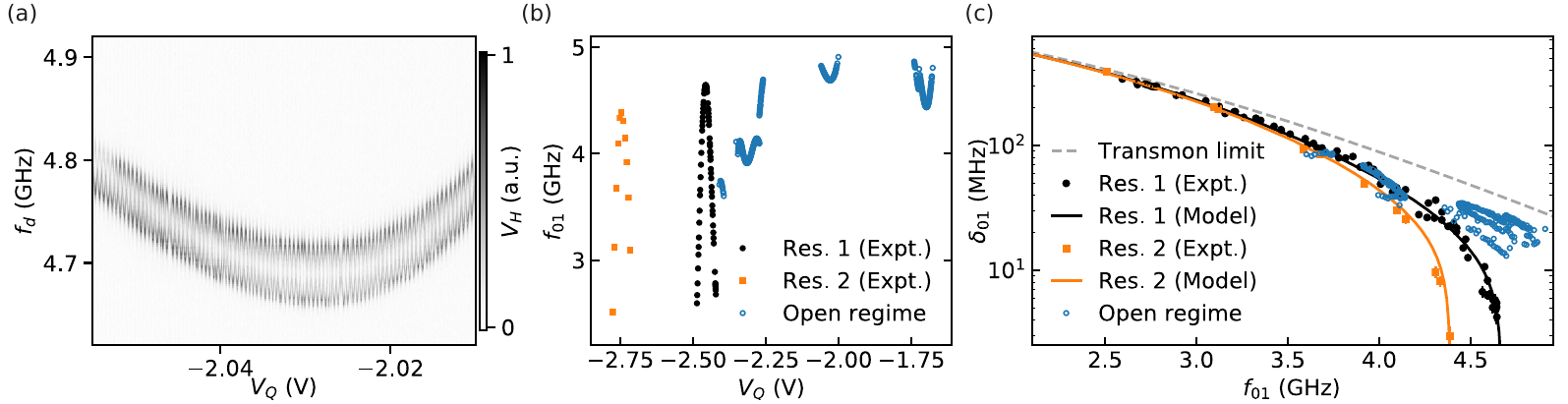}\vspace{-2mm}
    \caption{\vspace{-0mm} Charge dispersion when increasing $V_Q$. (a) An example of a measurement of $\delta_{01}$ in the open regime. (b) $f_{01}$ as a function of $V_Q$. The data points across Res.~1 (black) and Res.~2 (orange) are the same as presented in the inset of Fig.~4 in the main text. As $V_Q$ is increased further we extract $f_{01}$ in the open regime (blue data points). For $V_Q>-1.7$\,V we can no longer resolve $\delta_{01}$, but $f_{01}$ is still resolvable. (c) $\delta_{01}$ as a function of $f_{01}$. The data points across Res.~1 (black) and Res.~2 (orange) and theory curves are the same as presented in Fig.~4 in the main text for comparison to the measured $\delta_{01}$ in the open regime (blue). 
    }
    \label{fig:open regime}
\end{figure}

In the data presented in the main text we focus on the charge dispersion of the two dot resonances appearing near the pinch-off voltage of the qubit junction. We also extract the dispersion as $V_Q$ is increased. The dispersion is measured in the same way as for Res.~1, where $V_Q$ is swept finely to both vary $n_g$ and the qubit frequency [Fig.~\ref{fig:open regime}(a)]. In Figs.~\ref{fig:open regime}(b) and (c) we show dependence of $f_{01}$ on $V_Q$ and a parametric plot of the extracted $\delta_{01}$ values as a function of $f_{01}$, plotted together with the data and curves presented in Fig.~4 of the main text. Here we observe a deviation compared to the transmon dispersion. However, the suppression is not as extreme as observed for the resonances. We attribute this to the transmission not approaching unity as dramatically in this regime, but rather that transport across the junction is described by a few highly transmitting modes. We also observe a non-monotonic behaviour in both $f_{01}$ and $\delta_{01}$ as a function of $V_Q$. We interpret this as crossing from a resonant tunneling regime where narrow controlled resonances are observed to a regime where mesoscopic fluctuations in the nanowire junction results in an uncontrolled variation of individual transmission coefficients as a function of $V_Q$.

\end{document}